\documentclass[apj]{emulateapj}

\usepackage{natbib}
\usepackage{graphicx}
\bibpunct{(}{)}{;}{a}{}{,}

\newcommand\modified[1]{{#1}}

\begin{document}

   \title{
Qualitative interpretation of galaxy spectra
%
%
%
%
}

   \author{
          	J.~S\'anchez~Almeida\altaffilmark{1,2},
                R. Terlevich\altaffilmark{3,4},
                E. Terlevich\altaffilmark{3},
                R. Cid Fernandes\altaffilmark{5},
                \and
         	A.~B. Morales-Luis\altaffilmark{1,2}
          }
\altaffiltext{1}{Instituto de Astrof\'\i sica de Canarias, E-38205 La Laguna,
Tenerife, Spain}
\altaffiltext{2}{Departamento de Astrof\'\i sica, Universidad de La Laguna,
Tenerife, Spain}
\altaffiltext{3}{Instituto Nacional de Astrof\'\i sica, \'Optica y Electr\'onica, Tonantzintla, Puebla, Mexico}
\altaffiltext{4}{Institute of Astronomy, University of Cambrige, Cambridge, UK}
\altaffiltext{5}{Departamento de F\'\i sica - CFM - Universidade Federal de Santa Catarina, 
PO Box 476, 88040-900, Florianopolis, SC, Brazil}
\email{jos@iac.es,rjt@ast.cam.ac.uk,eterlevi@inaoep.mx, cid@astro.ufsc.br,abml@iac.es}
\begin{abstract}
We describe a simple step-by-step guide to qualitative 
interpretation of galaxy spectra (Fig.~\ref{decision}). Rather than
an alternative to existing automated tools, it is 
put forward as an instrument  for quick-look analysis, and for gaining 
physical insight when interpreting the outputs provided by automated tools. 
Though 
the recipe is of general application, it was  developed for 
understanding the nature of the Automatic Spectroscopic K-means 
based (ASK)  template spectra. They resulted from the  classification of all the galaxy 
spectra in the Sloan Digital Sky Survey data release 7 (SDSS-DR7), thus being a 
comprehensive representation of the galaxy spectra  in the local 
universe. Using the recipe, we give a description of the 
properties of the gas and the stars that characterize the
ASK classes, from those corresponding to passively evolving galaxies, 
to HII galaxies undergoing a galaxy-wide starburst. 
The qualitative analysis is found to be in excellent
agreement with quantitative analyses of the same spectra.
We compare the mean ages of the stellar populations with those
inferred using the code {\sc starlight}.
We also  examine the estimated gas-phase metallicity with the metallicities 
obtained using electron-temperature based methods.
A number of byproducts follow from the analysis. There is  a tight correlation 
between the age of the stellar population and the metallicity of the gas, which is stronger
than the correlations between galaxy mass and stellar age, and galaxy mass and gas 
metallicity. The galaxy spectra are known to follow a 1-dimensional sequence, 
and we identify the luminosity-weighted mean stellar age as the affine parameter that describes the sequence.
All ASK classes happen to have a significant fraction of old stars, although
spectrum-wise they are outshined by the youngest populations. 
Old stars are metal rich or metal poor depending on whether they reside
in passive galaxies or in star-forming galaxies. 
\end{abstract}


\slugcomment{}


%
%
\section{Introduction}\label{intro}


        There are several automated tools for inferring
the properties of the stellar populations contributing
to the integrated galaxy spectra. The list includes
{\sc moped} \citep{2004MNRAS.355..764P}, 
{\sc starlight} \citep{2005MNRAS.358..363C},
{\sc steckmap} \citep{2006MNRAS.365...74O},
{\sc vespa} \citep{2007MNRAS.381.1252T},
or {\sc ulyss} \citep{2009A&A...501.1269K},
as well as the use of line indices like the Lick 
indices \citep{1994ApJS...94..687W}.
Similarly,  there are semi-automatic 
procedures to deduce the properties
of the gas 
\citep[e.g.,][]{1995PASP..107..896S,2006IAUS..234..439J,pynebular},
including the so-called strong-line ratio methods
\citep[e.g.,][]{1979MNRAS.189...95P,2000MNRAS.312..130D,2002MNRAS.330...69D,2005A&A...437..849S}.
These tools are (and will be) fundamental for understanding the 
galaxy formation and evolution, but the blind use of 
the codes results quite unsatisfactory from a physical stand 
point. One obtains a  precise quantitative 
description of the stellar populations contributing to the 
integrated spectra, but ignores the reason why the code 
has chosen them rather than other potential alternatives. 
The educated
eye of an astronomer is often far more telling from a physical
point of view. Unfortunately, the know-how of qualitatively 
interpreting a spectrum is learned after a long experience
of working in the field. The information on which particular
spectral feature informs of which particular physical property
is scattered among a large number of technical publications,
difficult to identify and to deal with for a newcomer. This paper aims at 
providing
a step-by-step guide to qualitative interpretation of galaxy 
spectra. Moreover, it will be compared with up-to-date numerical techniques
to show that both qualitative and quantitative results are in excellent agreement.

The work was originally planned as a mere academic 
exercise to understand the nature of the 
classes resulting from the  k-means classification of  all 
the galaxy spectra in the  Sloan Digital Sky Survey  data 
release 7 
\citep[SDSS-DR7, ][]{2010ApJ...714..487S}.
We wanted to translate the spectral shapes into physical units like stellar ages and 
metallicities, so that this information can be used to tailor class-based searches 
\citep[e.g.,][]{2012A&A...540A.136A},
or when interpreting spectra \citep[e.g.,][]{sanchezjanssen2012}.  
However, the exercise
is of interest beyond the original scope.
The simple decision tree we use is  suitable to 
characterize any galaxy spectrum. We know of its  
generality because it allows to separate and 
characterize the 28 Automated  Spectroscopic 
K-means-based (ASK) classes \citep{2010ApJ...714..487S} which, 
by construction, are proxies that condense the properties of the some
one-million SDSS spectra  \citep[][]{2002AJ....123..485S,2009ApJS..182..543A}. 
The ASK class characterization represents a significant part of the
paper, that are discussed in detail as an illustration 
of the procedure. As we stress above, 
our qualitative analysis may have several other applications, e.g., 
(1) to gain physical insight when interpreting quantitative 
Star Formation Histories (SFHs) derived from modern automated 
tools,
(2) for quick-look galaxy  classification (not only
in the local universe, but also at moderate-high redshifts, since 
the Hubble expansion shifts the UV-visible spectrum to the near IR), 
(3) for interpreting noisy spectra where 
eyeball inspection is often better than detailed inversion,
(4) as  reference for identifying unusual galaxies, 
or (5) for educational purposes to develop physical
intuition.

The paper is organized as follows. 
Section~\ref{ask_class} introduces the ASK spectral classification of 
galaxy spectra whose templates serve as reference point.
Section~\ref{list_features} lists and discusses spectral features 
commonly used when interpreting galaxy spectra. 
They are employed to set up the recipe introduced 
in Sect.~\ref{decision_tree}, which is  
abridged  in a schematic shown in Fig.~\ref{decision}.
The recipe (or algorithm) is used in Sect.~\ref{qualitative_classes}
to disclose the physical properties of all the ASK classes.
The results of such qualitative analysis are compared with 
state-of-the-art quantitative analyses in Sects.~\ref{starlight} and 
\ref{quantitative_lines} -- Sect.~\ref{starlight} deals with the comparison
of stellar components, whereas Sect.~\ref{quantitative_lines} refers
to the gas components. 
Section~\ref{additional_results} discusses several
additional properties of the ASK templates, whereas
Sect.~\ref{conclusions} summarizes the proposed qualitative
analysis.

\section{ASK classification}\label{ask_class}

\citet{2010ApJ...714..487S} classified all the galaxies with spectra in 
SDSS-DR7 into only 28 ASK classes. 
The original SDSS covers one-fourth of the sky and contains
the spectra of all the galaxies above 
an apparent magnitude threshold (SDSS~$r < 17.8$). Therefore,
the some one million SDSS spectra can be regarded as representative of the 
galaxies of the local universe,  and so do the ASK classes
inferred from them. The ASK classification is detailed in 
\citet{2010ApJ...714..487S}, with additional properties of the 
classes discussed elsewhere 
\citep[][]{2011ApJ...735..125S,
2011MNRAS.415.2417A,
2012A&A...540A.136A}. 
For the sake of comprehensiveness, however, we 
summarize here the main properties. 
All galaxies with redshift smaller than 
0.25 were transformed to a common rest-frame 
wavelength scale, and then re-normalized to
the integrated flux in the SDSS $g$-filter. 
These two are the only manipulations the
spectra underwent before classification. 
We wanted the classification to be driven only 
by the shape of the visible spectrum
(from 400 to 770 nm), and these two 
corrections remove obvious undesired dependencies 
of the observed spectra on redshift and galaxy apparent
magnitude. We deliberately avoided correcting for other 
effects requiring modeling and assumptions 
(e.g., dust extinction, seeing, or aperture effects). 
%
%
The employed classification algorithm, k-means, is a robust workhorse
that allows the simultaneous classification of the
full data set ($\sim$12\,GB). It is commonly employed in data mining, machine learning, 
and artificial intelligence \citep[e.g.,][]{eve95,bis06}, 
and it guarantees that similar rest-frame spectra belong to 
the same class.
Most galaxies (99\,\%) were  assigned to only 17 major 
classes, with 11 additional minor classes including the
remaining 1\,\%.
It is unclear whether the ASK classes represent
genuine clusters in the 1637-dimensional classification
space, or if they 
\modified{
slice
}
a continuous distribution -- probably 
the two kinds of classes are present
\citep[see][]{2010ApJ...714..487S,
2011MNRAS.415.2417A}.

All the galaxies in a class have very similar spectra, which are 
also similar to the class template spectrum formed as the 
average of all the  spectra of the galaxies in the class. These template
spectra are the ones analyzed in the paper. The averaging is 
slightly different from the one in \citet[][]{2010ApJ...714..487S},
and the novelty allows us to reach the near UV of the spectrum. 
The SDSS spectrograph detects from 3800\,\AA\ to 9200\,\AA\
\citep[e.g.,][]{2002AJ....123..485S}, however, the templates cover 
from 3000\,\AA\ to  9200\,\AA . The UV extension is recovered 
because  the classified  galaxies have redshifts up to 0.25, which 
moves the rest-frame  
$\lambda\,$3000\,\AA\ within the observed range.   Rather 
than averaging the spectral range common to all 
galaxies, the new templates consider the full range
of available rest-frame wavelengths.
Given a wavelength bin, it includes the spectra of all the galaxies 
in the class that have been observed at that particular
rest-frame wavelength. Consequently, the
template spectra (i.e., the average spectra) include
wavelengths down to 3000\,\AA .
The templates thus obtained vary smoothly and 
continuously. They are labeled according to the $u-g$ color, 
from the reddest, ASK~0, to the bluest, 
ASK~27. The use of numbers to label the classes does not
implicitly assumes the spectra to follow
a one dimensional family. The numbers only
name the classes. The sorting (and, so, the naming) 
would have been slightly different using other bandpasses
to define colors. In general, however, the smaller the ASK class number
the redder the spectrum. 
The ASK classification  of all galaxies with spectra in SDSS-DR7 is publicly 
available, templates included\footnote{\label{my_foot}{\tt ftp://ask:galaxy@ftp.iac.es/}\\
{\tt http://sdc.cab.inta-csic.es/ask/index.jsp} in the 
Spanish Virtual Observatory.
}.
Wavelengths of SDSS spectra (and of ASK templates) are vacuum wavelengths. 
However, all the spectra shown in 
this paper are transformed to air 
wavelengths according to the equations by \citet{1996ApOpt..35.1566C}.
\modified{
The SDSS spectra used for classification, and so the templates 
shown along the paper, are given as flux per unit wavelength.
}

\section{Recipe for qualitative interpretation of galaxy spectra}
Galaxies have composite spectra. They integrate contributions from
different stars of different stellar populations, from HII regions, from Active Galactic
Nuclei (AGNs),
as well as from other possible components 
\citep[e.g., hydrogen ionized by old 
high temperature stars, or by the intergalactic UV background -- ][]
{2004MNRAS.355..273C,
2010MNRAS.403.1036C}.
Our qualitative analysis builds on this fact, and tries to separate 
each spectrum into a minimum number of components.   
We consider the (ionized) gas and the stars separately, that is
to say, the emission and absorption lines separately.
Each one of these two components is assumed to have 
one or two sub-components. The details on the 
characterization are summarized as a decision tree
in Sect.~\ref{decision_tree}. It is  based on the analysis of 
a set of general spectral features, listed in the next sub-section.

\subsection{Spectral features to be considered in a qualitative analysis}\label{list_features}
The main spectral features that can be considered are listed
in the section, ordered from the more obvious to the subtle details.
Each item names the particular feature, and then  outlines
its main properties and interest. The actual features are illustrated 
using the appropriate ASK templates.

\begin{enumerate}
\item The shape of the continuum and the presence or not of emission and absorption
lines must be considered. The emission lines trace the ionized
gas and its excitation mechanism. The absorption lines trace the stellar populations,
their ages and metallicities.
The overall continuum shape is modulated  by the gas, the stars, 
as well as by the  presence of dust.
Figure~\ref{dust_effect} shows the prototype 
red galaxy with passively evolving  stellar populations (ASK~0).  Although red, 
the continuum is rather flat  from 6000\,\AA\  on. Spectra 
even redder
must be shaped by dust extinction (see ASK~1 in Fig.~\ref{dust_effect}).  
\begin{figure}
\includegraphics[width=.5\textwidth]{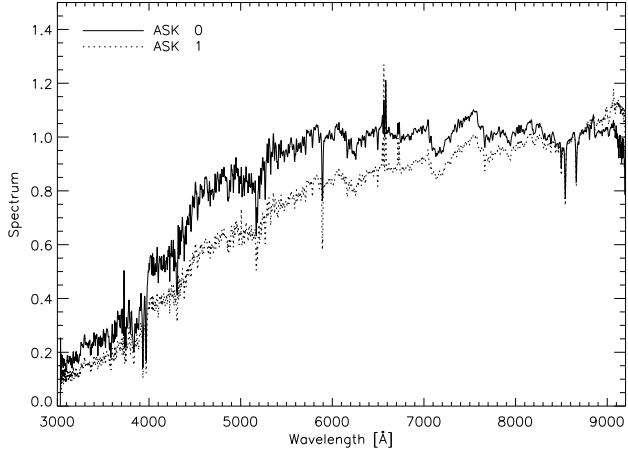}
\caption{
Effect of dust in the continuum. ASK~0 is the prototype of galaxy with
red stellar populations. Since ASK~1 is even redder (i.e., the continuum
has larger slope), its continuum must be affected by significant extinction.
The presence of strong emission lines in ASK~1 reinforces the conclusion.   
\modified{
The spectra are given as flux per unit wavelength, here and throughout
the paper.
}
}
\label{dust_effect}
\end{figure}
\item The so-called 4000\,\AA~{\em break} is produced by the absorption
of metallic lines of a  variety of elements in various states of ionization,
including Ca\,II~H~and~K ($\lambda\lambda$\,3969\,\AA\ and 3934\,\AA ) and
high-order lines of the Balmer series 
(H$\epsilon\lambda$\,3970\,\AA, 
H$\zeta\lambda$\,3889\,\AA	, 
H$\eta\lambda$3835\,\AA, \dots) 
\citep[see ][ and also Fig.~\ref{uvbreaks}]{1985ApJ...297..371H}.  
The opacity suddenly increases for photons bluer than this wavelength, 
which  produces an intensity drop. 
It is enhanced in old stellar populations (ASK~0), which tend
to be metal rich, but it is also present in younger 
galaxies (ASK~19 in Fig.~\ref{uvbreaks}).
The Balmer lines become deeper and broader
with time from the starburst, with a characteristic time-scale of the order of one~Gyr 
\citep[e.g.,][]{1999ApJS..125..489G}.
\begin{figure}
\includegraphics[width=.5\textwidth]{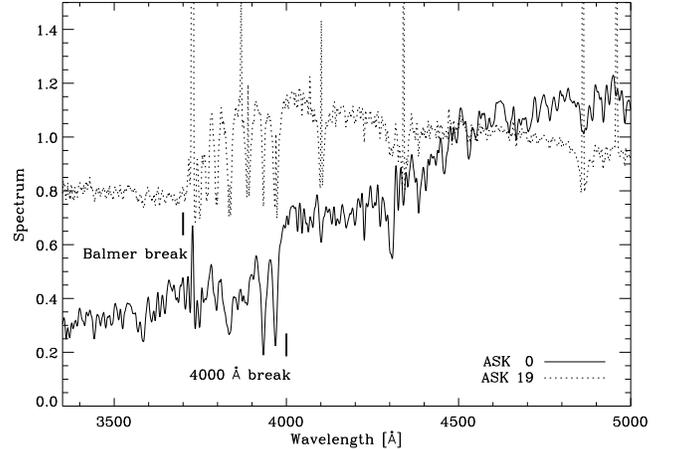}
\caption{
Spectra showing the 4000\,\AA~break , as well as the
break produced by the  Balmer limit at 3650\,\AA . 
}
\label{uvbreaks}
\end{figure}
\item The limit of the Balmer series and the blending of the high-order
Balmer lines
also produces a notable discontinuity of the spectrum
blueward of
3650\,\AA . It is the {\em Balmer break} --
see Fig.~\ref{uvbreaks}. (Photons bluer than 
this limit ionize the excited
hydrogen, thus H becomes an important source of
continuum opacity.) It is present in young and old stellar
populations, but it is more important in the young populations
where H is a major constituent of the opacity (especially
in the Balmer continuum beyond the discontinuity). The
break amplitude and position is a proxy for the age of the stellar population 
\citep[e.g.,][]{2001MNRAS.325..636A}.
\item 
The Ca\,II~H and K lines ($\lambda\lambda$  3969\,\AA\ and 3934\,\AA , respectively) 
are typical of old metal-rich stars. 
Ca\,II~H
is blended with H$\epsilon$ which, 
as the rest of the Balmer series, appears in absorption in 
young stars (say, A~stars). 
In case of mixed populations
of old and young stars, the relative intensities of 
Ca\,II~H and Ca\,II~K (actually, of Ca\,II~K and Ca\,II~H+H$\epsilon$)
is a proxy for the relative importance of the young and old populations.
When Ca\,II~K is larger than Ca\,II~H,  then the old population dominates
the spectrum (ASK~2 in Fig.~\ref{cahk}). As the young population becomes 
more important then Ca\,II~H becomes 
stronger than Ca\,II~K 
(ASK~9 in Fig~\ref{cahk}). 
The relative growth reverts when the HII regions accompanying the young stellar
populations  produce enough  H$\epsilon$ emission, which fills the 
Ca\,II~H+H$\epsilon$ absorption profile (ASK~14  in Fig~\ref{cahk}).  
 \begin{figure}
\includegraphics[width=.5\textwidth]{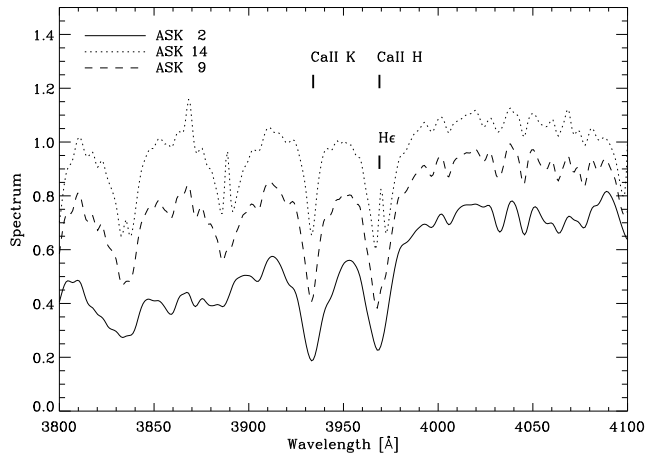}
\caption{
Spectra in the range of the
Ca\,II~H and K lines.
 Ca\,II~H is blended with H$\epsilon$. In case of mixed young and old stellar
populations, the relative importance of Ca\,II~K and the blended Ca\,II~H 
informs on the relative importance of the two populations. 
Ca\,II~K dominates in old populations (ASK~2), and  Ca\,II~H
dominates when the young stars are more important (ASK~9), but
the relationship saturates when H$\epsilon$ starts to show up in emission (ASK~14). 
}
\label{cahk}
\end{figure}  
\item The UV continuum flux is also an age indicator for very young 
stellar populations.
It increases with decreasing age when the ages are only a few Myr --
see Fig.~\ref{uvcontinuum} and, e.g., \citet{1999A&A...349..765M}.
A symptom of extreme  youth is the Balmer continuum showing up in emission 
($\lambda < 3650$\,\AA ), as it happens with ASK~25 in Fig.~\ref{uvcontinuum}.
\begin{figure}
\includegraphics[width=.5\textwidth]{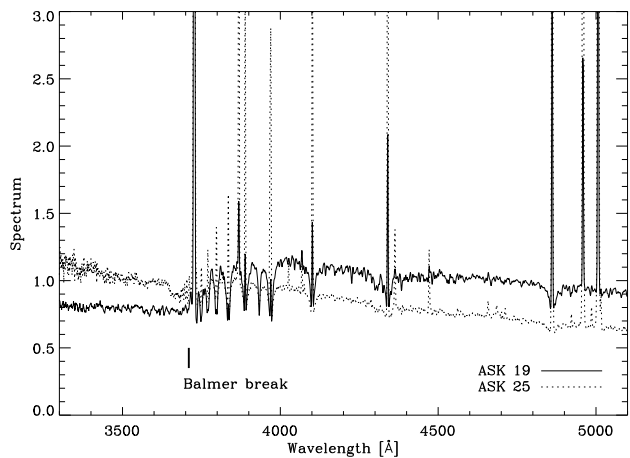}
\caption{
Spectra showing the increase of the UV continuum to the blue of 3650\,\AA\
as the stellar populations become  younger (from ASK~19 to ASK~25).
}
\label{uvcontinuum}
\end{figure}
\item The ratio between the fluxes of H$\alpha$ and 
[NII]$\lambda$6583 is an indicator of whether the 
nebula is ionized by a starburst   (H$\alpha \gg  {\rm [NII]}\lambda6853$),
or by a source of harder UV flux like an AGN or 
low mass  evolved stars 
(${\rm H}\alpha/2.5 \le {\rm [NII]}\lambda6853$).
Figure~\ref{star_agn} shows examples of the two
possibilities: a starburst (ASK~9) and a LINER-like 
excitation\footnote{We use the term {\em LINER-like} to refer 
to real
Low-ionization nuclear emission-line regions \citep{1980A&A....87..152H}, 
or evolved stars in retired galaxies \citep[][]{2011MNRAS.415.2182F}, or
 X-ray emitting gas \citep[e.g.,][]{2012ApJ...747...61Y},
or any other source with an ionizing UV spectrum harder 
than that of newborn stars, but not as hard as 
in a Seyfert galaxy or a quasar.
\modified{
Shock heated gas may also produce lines in this part
of the diagram \citep[e.g.,][]{1981PASP...93....5B,2008ApJS..178...20A}.
}
}
(ASK~0). Once the ratio [NII]$\lambda$6583 to  H$\alpha$ is 
observed to be outside the starburst regime, 
Seyferts  and  LINERs can be distinguished according to  
the ratio of fluxes between [OIII]$\lambda$5007 and H$\beta$ 
(Fig.~\ref{bpt2nd}) -- the high ionization source is a  Seyfert
if [OIII]$\lambda 5007 \gg {\rm H}\beta$ (ASK~6) or 
LINER-like  if [OIII]$\lambda$5007 $\lesssim$ H$\beta$ (ASK~5). 
This recipe is a qualitative rendering of the so-called
BPT diagram  \citep{1981PASP...93....5B} and its 
updates \citep{2003MNRAS.346.1055K,2006MNRAS.372..961K,2010MNRAS.403.1036C}.
\begin{figure}
\includegraphics[width=.5\textwidth]{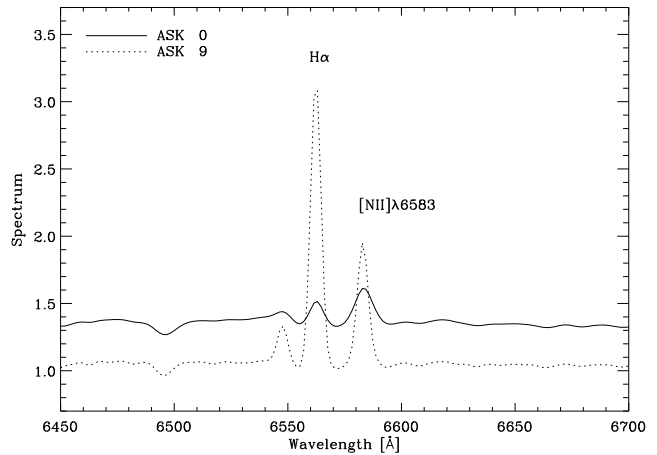}
\caption{
The ratio between H$\alpha$ and 
[NII]$\lambda$6583 indicates whether the nebula is ionized
by a starburst, or by other type of source
with higher ionization power (e.g., AGNs).
H$\alpha \gg {\rm [NII]}\lambda6853$ indicates
starburst (ASK~9) whereas ${\rm [NII]}\lambda6853 \ge {\rm H}\alpha$
is a symptom of higher ionization (ASK~0).
}
\label{star_agn}
\end{figure}
\begin{figure}
\includegraphics[width=.5\textwidth]{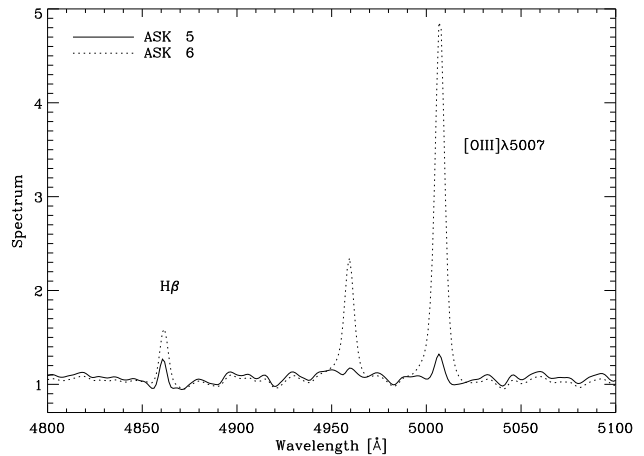}
\caption{
After establishing that the galaxy is not a starburst,
the ratio between H$\beta$ and 
[OIII]$\lambda$5007 indicates whether the
ionization is powered by a strong AGN 
([OIII]$\lambda$5007 $\gg$ H$\beta$ -- ASK~6)
or by a LINER-like  source ([OIII]$\lambda$5007 $\lesssim$ H$\beta$ -- ASK~5).
}
\label{bpt2nd}
\end{figure}
%
\item \label{bptitem}  The ratio between the fluxes of [NII]$\lambda$6583 
and  H$\alpha$ also provides an estimate of gas
metallicity in star-forming galaxies \citep{2002MNRAS.330...69D}.
The sensitivity is high: the ratio goes from 1/3 to 1/300
when  the metallicity ranges from solar to one tenth solar
\citep{2004MNRAS.348L..59P}. This high contrast makes it simple to 
distinguish between solar and sub-solar metallicities. 
\modified{
The ratio is not suitable to diagnose super-solar metallicities.
In this case one can use supplementary line ratios  
like the so-called R23 or 03N2 \citep[e.g.,][]{1979A&A....78..200A,
1979MNRAS.189...95P,2005A&A...437..849S}. 
}
%
\item \label{line4363} The presence of [OIII]$\lambda$4363 is also an indicator of low 
metallicity. The line is used to compute electron temperatures
in HII regions, and it weakens with increasing metallicity 
to disappear at around $12+\log({\rm O/H})\gtrsim 8.2$ 
\citep[e.g.,][]{1991ApJ...380..140M}.
\item  TiO bands at 
approximately  7150\,\AA , 7600\,\AA , and 8500\,\AA\ 
are characteristic of M stars 
\citep[dwarf, giant and super-giants; e.g.,][]{2000ApJ...540.1005A},
and reveal the presence of evolved stellar populations.
When the stellar population is young, massive stars outshine the 
contribution of M stars, making these spectral features 
invisible.
\modified{ 
(Red super-giants are young massive stars showing TiO bands, but
they are outnumbered  by the associated blue super-giants that
overshadow their contribution to the integrated galaxy spectrum
-- e.g., \citeauthor{1980A&A....90L..17M}~\citeyear{1980A&A....90L..17M}; 
\citeauthor{2002ApJS..141...81M}~\citeyear{2002ApJS..141...81M}.)
}
Figure~\ref{tiobands} shows how the TiO bands 
appear in almost all ASK classes, except for the 
bluest ones, and how the strength of the bands 
decrease as the spectrum becomes bluer. 
\modified{
As we mention in Sect.~\ref{qualitative_classes}, 
the bands are hardly noticeable at ASK~16, and 
they are absent at ASK~20 and bluer types.
}
\begin{figure}
\includegraphics[width=.5\textwidth]{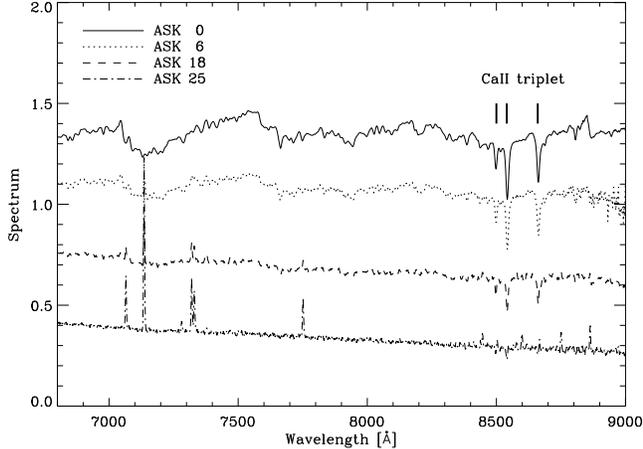}
\caption{
The TiO bands 
at approximately 7150\,\AA , 7600\,\AA , and 8500\,\AA\ 
are characteristic of M  stars, and they 
appear in all ASK classes except for the bluest ones
(ASK~25 in the plot).
Note also the presence of the IR~Ca\,II~triplet in the middle of the 
third TiO band ($\lambda\lambda$8498, 8542, and 8662\,\AA).
}
\label{tiobands}
\end{figure}
\item The IR~Ca\,II~triplet  at $\lambda\lambda$8498, 8542, and 8662\,\AA\
is an indicator of metallicity and gravity. In stars, its equivalent width (EW)
increases  with  increasing metallicity until 2/3 of the solar 
metallicity \citep{1989MNRAS.239..325D}. 
\modified{
Above this metallicity 
it depends only on gravity, with the EW 
increasing with decreasing gravity from dwarfs to super-giants
\citep{1989MNRAS.239..325D,2002MNRAS.329..863C}.
}
The combined effect on galaxy spectra must be modeled,
but the existence of a Ca\,II absorption with significant 
strength is always a sign of high 
metallicity and of the presence of giant stars. Contrarily,  
absence of the triplet indicates low metallicity.
We find it  in all ASK spectra except for the 
bluest classes (see Fig.~\ref{tiobands}).
\modified{
As we mention in Sect.~\ref{qualitative_classes}, 
the lines are almost absent in ASK~20 and bluer classes.
}
\item The so-called Mg$_2$ 
and H$\beta$ Lick indices are
in the same spectral region (H$\beta$ from 4848 to 4877\,\AA , and
Mg$_2$ from 5154 to 5197\,\AA ; see Fig.~\ref{hbetamg}), and 
they were designed (and are used) to determine simultaneously 
age and metallicity in galaxies with old stellar populations
\citep{1994ApJS...94..687W}.
\modified{
One 
}
can  generally say that H$\beta$ mostly depends on age, and to less extent
on metallicity, and the opposite happens with Mg$_2$ 
\citep[e.g.,][]{1996ApJS..106..307V,1999MNRAS.306..607J}.
\modified{
However, their quantitative application require
modeling 
(e.g., they may depend on the relative abundance of the metals,
rather than on a single global metallicity).
}
\begin{figure}
\includegraphics[width=.5\textwidth]{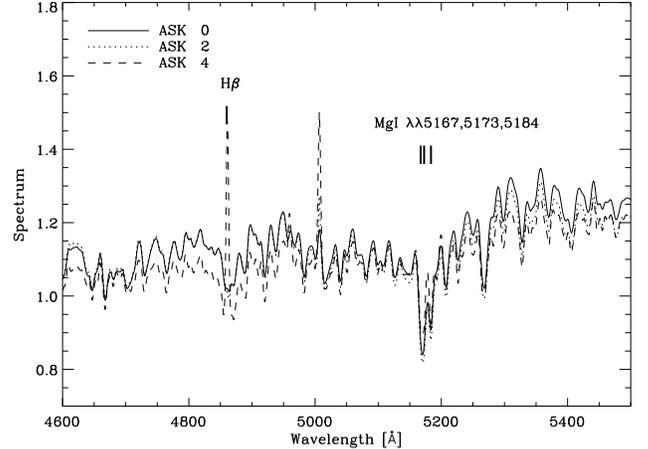}
\caption{
Spectral region containing the spectral
indices H$\beta$ and Mg$_2$  (H$\beta$ index from 4848 to 4877\,\AA , 
and Mg$_2$ from 5154 to 5197\,\AA). Both indices
combined allow us to set mean age and mean
metallicity in galaxies with old stellar populations.
The labels mark the H$\beta$ line and the position of the
three Mg~I lines contributing to Mg$_2$.
}
\label{hbetamg}
\end{figure}

\item The interstellar medium (ISM) that reddens the  
spectra also produces absorption in 
the Na\,I~D line 
\citep[$\lambda\lambda$5891,5896\,\AA , e.g.,][]{2007MNRAS.381..263A,2010AJ....140..445C}.
Therefore, one would 
expect that the strength of the ISM Na\,I~D line sorts
galaxies according to extinction \citep{2007MNRAS.381..263A}. The example in Fig.~\ref{nadline}
corresponds to the two spectra in  Fig.~\ref{dust_effect}, where 
ASK~1 is known to present  a substantial dust extinction. Its 
Na\,I~D  is  stronger than that for the class without extinction,
ASK~0, being the rest of the spectrum similar.
%
\begin{figure}
\includegraphics[width=.5\textwidth]{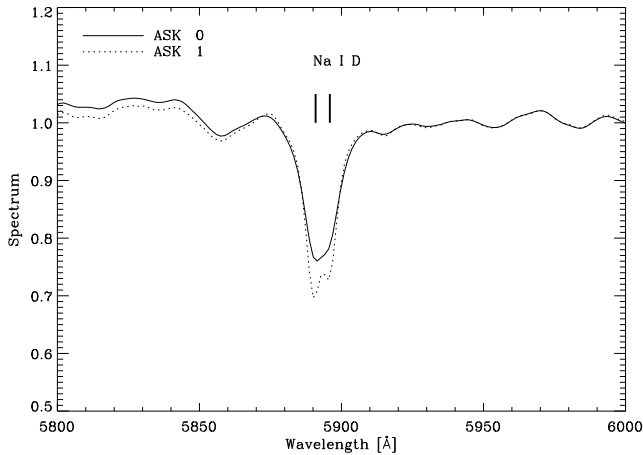}
\caption{
 Na\,I\,D is partly produced by ISM absorption.
It is larger for the class with larger dust extinction 
(c.f., Fig.~\ref{dust_effect}).
}
\label{nadline}
\end{figure}
\item The mere presence of high excitation lines like 
[Ne\,V]$\lambda$3426,
[Fe\,VII]$\lambda$6087,
or  [Fe\,X]$\lambda$6375,
tells us that the galaxy hosts an AGN
\citep[e.g.,][]{2003MNRAS.343..192R,2009MNRAS.398.1165G,2011ApJ...743..100R}.
The example in Fig.~\ref{agn_excitation}  
shows spectra in the range of  [Fe\,VII]$\lambda$6087, 
whose emission is clear in
Seyferts (ASK~7 and 8), but is
non-existing in starbursts (ASK~20) as well as in 
passively evolving red 
galaxies with LINER-like emission (ASK~0; see Sect.~\ref{qualitative_classes}).
\modified{
[He\,II]$\lambda$4686 is also indicative of AGN, though it is sometimes 
found in star forming galaxies
\citep[e.g.,][]{1999A&AS..136...35S,2012MNRAS.421.1043S}.
}
\begin{figure}
\includegraphics[width=.5\textwidth]{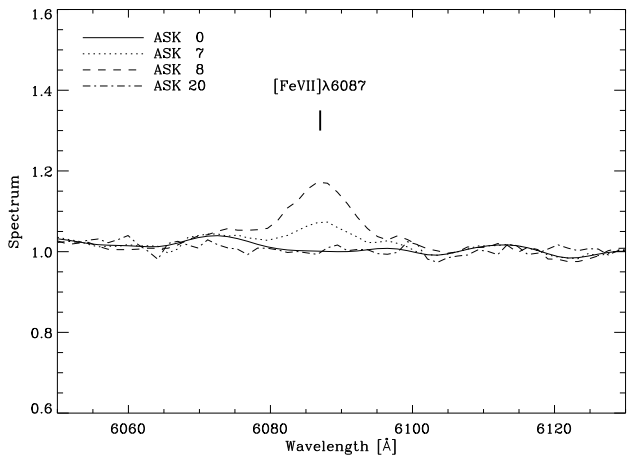}
\caption{
Spectral range containing the high excitation line
 [FeVII]$\lambda$6087, clear in 
Seyferts (ASK~7 and 8) but  non-existing otherwise.  
(ASK~20 is a  starbursts, whereas
ASK~0 represents a passively evolving red galaxy.)
}
\label{agn_excitation}
\end{figure}
\end{enumerate}

\subsection{Decision tree for qualitative analysis of galaxy spectra}\label{decision_tree}
\begin{figure*}
\includegraphics[width=0.8\textwidth]{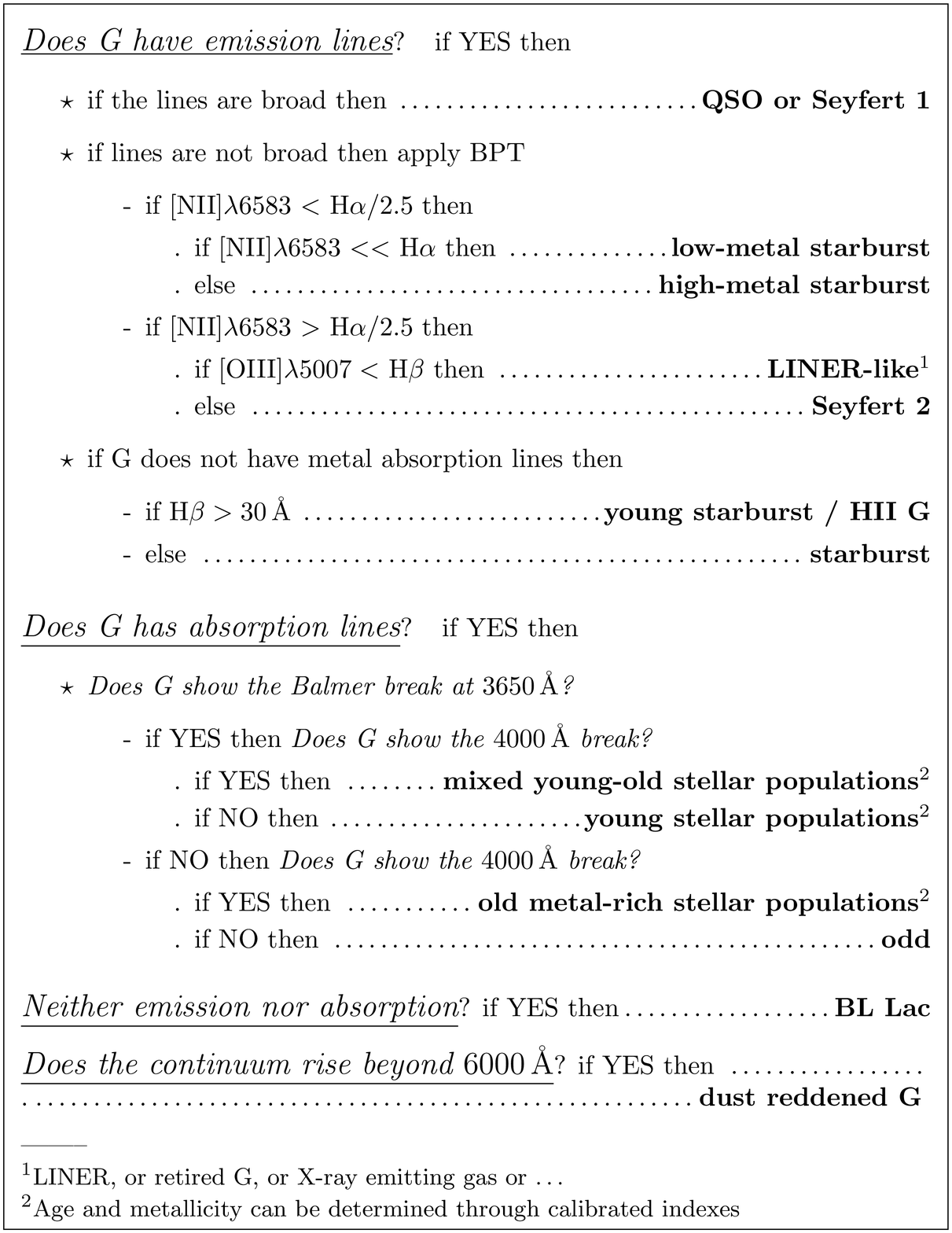}
\caption{Decision tree for  qualitative classification of galaxy spectra.
The questions to be replied are written in italic fonts.
The galaxy types resulting from the 
analysis are written in bold fonts on the right hand side of the panel. 
The symbol G stands for galaxy.
See Sect.~\ref{decision_tree} for further details.}
\label{decision}
\end{figure*}

Considering the spectral features described in the previous section, 
we have set up a simple decision tree  (a questionary)
that leads to  classifying a galaxy spectrum by replying  to a few 
questions (Fig.~\ref{decision}). 
Emission and absorption lines are analyzed
separately, therefore, the natural outcome would be galaxy
types with two components, one for the stars and other
for the gas. 
One should begin the questionary from top to bottom, to
end up with the characteristics of both the gas and the stars. 

The decision tree in Fig.~\ref{decision} is self-explanatory, although 
a few clarifications on the terminology are required.
The symbol G stands for galaxy.
{\em Broad spectral lines} means lines in excess of 2000~km\,s$^{-1}$, and 
they separate Seyfert 1 and quasars from the other kinds
of AGNs. Such broad lines are not illustrated in Sect.~\ref{list_features}
since the ASK classes lack Seyfert~1 and quasars, that
were excluded from the  list of galaxy targets directly at the SDSS 
distribution \citep[see][]{2007AJ....134..102S}.  
When we mention young, old and a mixture of old-and-young stellar
populations, we loosely speaking refer to stellar ages $< 10^7$\,yr (young),
$>10^9$\,yr (old), and the intermediate range in between, $10^7$--$10^9$\,yr.
When metal poor gas is mentioned, we mean clearly sub-solar
(say, less than 1/3 solar).
BL Lac objects are also included to complete the questionary, so that
it considers the possibility that neither emission nor absorption lines
are present in the spectrum \citep[][]{1976ARA&A..14..173S,2012MNRAS.422.2322M}.
Several criteria in Fig.~\ref{decision}
compare emission lines -- such comparison refers to the fluxes
of the lines. 

%
%
\section{Qualitative analysis of the ASK classes}\label{qualitative_classes}

One may think of this section as analogous to the section on {\em individual objects},
common in many papers, 
except that the targets are ASK templates representing objects too 
numerous to be described one by one.
We use the criteria put forward in the previous section to determine 
the properties  of all ASK classes individually.
The thread of the argumentation  follows the decision tree in Fig.~\ref{decision}. 
A summary with the properties of all classes is also given in Table~\ref{table_summary}.
At the end of the section, we define a stellar age index that sorts the 
ASK classes according to their mean 
stellar age. Similarly, we define an index
to sort the emission line spectra  by metallicity. Both are relative quantities,
devised to compare our qualitative analysis with quantitative estimates
of ages and metallicities.
\smallskip

\setlength{\tabcolsep}{0.02in} 
\begin{deluxetable*}{clllccccc}
\tabletypesize{\scriptsize}
\tablecolumns{9}
\tablewidth{0pt}
\tablecaption{Qualitative physical properties of the ASK template spectra}
\tablehead{
\colhead{ASK}&
\colhead{emission}&	
\colhead{absorption}&
\colhead{comment}&
\colhead{SAI\tablenotemark{c}}&
\colhead{GMI\tablenotemark{d}}&
\colhead{O/H\tablenotemark{e}}&
\colhead{Age [Gyr]\tablenotemark{f}}&
\colhead{$Z/Z_\odot$\tablenotemark{f}}\\
\colhead{class}&
\colhead{(gas)}&
\colhead{(stars)}&
&
&
&
\colhead{(gas)}&
\colhead{(stars)}&
\colhead{(stars)}
}
\startdata
0&AGN or LINER-like&old metal-rich~~& H$\alpha$ EW$\simeq$0.9\AA \tablenotemark{a}&0&---&---&11.2$\pm$1.2&1.30$\pm$0.34\\
1&LINER-like& old metal-rich& dust reddened, edge on disks?\tablenotemark{b}&0&---&---&10.0$\pm$1.4&0.91$\pm$0.34\\
2&AGN or LINER-like &old metal-rich& [SIII]$\lambda$9069 emission&1&---&---&11.1$\pm$1.2&1.31$\pm$0.34\\
3&LINER-like	&old metal-rich& continuum bluer than ASK~0 and 2&2&---&---& 6.7$\pm$1.2&1.57$\pm$0.38\\
4&LINER-like?&old \& young&edge-on disks?\tablenotemark{b}&3&---&---& 7.4$\pm$1.5&0.69$\pm$0.36\\
5&LINER-like?&old \& young&green valley galaxies\tablenotemark{a}&3&---&---& 6.0$\pm$1.4&1.21$\pm$0.43\\
6&Seyfert 2&old \& young&[FeVII]$\lambda$6087 emission&3&---&---& 5.3$\pm$1.3&1.35$\pm$0.44\\
7&Seyfert 2&old \& young& younger than 6, [FeVII]$\lambda$6087 emission&4&---&---& 5.2$\pm$1.2&1.23$\pm$0.44\\
8&Seyfert 2&old \& young& younger than 7, [FeVII]$\lambda$6087 emission&5&---&---& 2.30$\pm$0.71&1.05$\pm$0.44\\
9& LINER-like&old \& young& metal-rich starburst?&3&---&---&3.6$\pm$1.2&1.01$\pm$0.44\\
10&metal-rich starburst&old \& young&LINER-like?&3&-0.35&---&4.1$\pm$1.3&0.61$\pm$0.34\\
11&metal-rich starburst&old \& young&LINER-like?, stars younger than 9 and 10&4&-0.36&---& 4.8$\pm$1.4&0.43$\pm$0.24\\
12&metal-rich starburst&old \& young&starburst prototype, stars younger than 11&6&-0.43&8.46$\pm$0.18& 2.7$\pm$1.1&0.68$\pm$0.35\\
13&metal-rich starburst&old \& young&stars similar to 12&6&-0.46&--- &2.30$\pm$0.93&0.90$\pm$0.41\\
14&metal-rich starburst&old \& young&starburst prototype, stars younger than 12&7&-0.46&8.50$\pm$0.11&1.71$\pm$0.92&0.60$\pm$0.30\\
15&metal-poor starburst&no absorption & HII G, youngest ASK&--&-1.67&7.85$\pm$0.05&---&---\\
16&metal-poor starburst&old \& young&---&7&-0.63&8.77$\pm$0.10& 1.18$\pm$0.72&0.58$\pm$0.30\\
17&metal-poor starburst& young& HII G, stars older than 15&13&-1.58&8.06$\pm$0.02& 0.0048$\pm$0.0008&1.29$\pm$0.43\\
18&metal-poor starburst& young& stars younger than 16&8&-0.54&8.61$\pm$0.04&0.090$\pm$0.038&0.52$\pm$0.24\\
19&metal-poor starburst& young& stars as in 18&8&-0.75&8.72$\pm$0.05& 0.249$\pm$0.077&0.51$\pm$0.25\\
20& metal-poor starburst&young& HII G, stars younger than 18, older than 17&13&-1.42&8.19$\pm$0.01&0.0045$\pm$0.0007&0.60$\pm$0.33\\
21& metal-poor starburst&young& HII G, like 20, gas slightly metal-poorer&13&-1.45&8.07$\pm$0.01& 0.0073$\pm$0.0019&0.72$\pm$0.34\\
22& metal-poor starburst&young& like 19, stars younger, gas metal-richer&8&-0.93&8.59$\pm$0.04& 0.138$\pm$0.057&0.39$\pm$0.17\\
23& metal-poor starburst&young& like 19 and 22, stars younger&9&-0.79&8.60$\pm$0.02& 0.056$\pm$0.031&0.45$\pm$0.24\\
24& metal-poor starburst&young& like 23, stars younger&10&-1.07&8.48$\pm$0.02 & 0.062$\pm$0.035&0.44$\pm$0.22\\
25& metal-poor starburst&young& like 20 and 21, stars older, gas metal-richer&12&-1.27&8.29$\pm$0.01& 0.0083$\pm$0.0018&0.72$\pm$0.36\\
26& metal-poor starburst&young& like 25, stars older, gas metal-richer&11&-1.09&8.38$\pm$0.02 & 0.0090$\pm$0.0020&0.49$\pm$0.28\\
27& metal-poor starburst&young& like 25&12&-1.18&8.23$\pm$0.02& 0.0098$\pm$0.0033&0.71$\pm$0.34
\enddata
\tablenotetext{a}{From \citet{2010ApJ...714..487S}, Table~2.}
\tablenotetext{b}{\citet{2011ApJ...735..125S}.}
\tablenotetext{c}{Stellar Age Inex, which orders stellar ages staring from the oldest (SAI=0)}
\tablenotetext{d}{Gas Metallicity Index
($\equiv\log({\rm [NII]}\lambda 6853/{\rm H}\alpha)$), which orders 
galaxies according to gas metallicity.}
\tablenotetext{e}{$12+\log({\rm O/H})$ obtained via electron temperature and density. The error bars 
account for uncertainties inherited from errors in line fluxes.}
\tablenotetext{f}{Luminosity-weighted average values using {\sc starlight}. The error bars represent
the dispersion among the SSP that contribute to the integrated light.}
\label{table_summary}
\end{deluxetable*}

$\bullet$ ASK~0 has an absorption line spectrum with very weak emission lines
(Fig.~\ref{dust_effect}). It is not a starburst since  ${\rm [NII]}\lambda6853 > {\rm H}\alpha$,
but [OIII]$\lambda$5007 and H$\beta$ are too weak to decide whether the excitation
is Seyfert-like or LINER-like.  Note, however, that the EW of H$\alpha$ is 
very small (Fig.~\ref{star_agn}),  which according to \citet{2011MNRAS.413.1687C}
indicates that the ionization is produced by hot-low mass stars.  
The absorption line spectrum does not show the Balmer break (Fig.~\ref{uvbreaks})
but the 4000\,\AA\ break is conspicuous, consequently, the absorption
spectrum is produced by an old metal-rich stellar population.

{$\bullet$ ASK~1} also has an absorption line spectrum with 
weak  emission lines (Fig.~\ref{dust_effect}). 
${\rm [NII]}\lambda6853 \simeq {\rm H}\alpha$, and therefore it
is not a starburst. H$\beta$ is smaller than 
 [OIII]$\lambda$5007, which may naively indicate AGN excitation. However,
the lines are so weak that the underlying H$\beta$ absorption is 
important and, therefore the corrected H$\beta$ emission 
is similar to that of  [OIII]$\lambda$5007. Consequently, the emission line spectrum is 
probably in the LINER region of the BPT diagram. The absorption line spectrum is also very 
similar to ASK~0, which was assigned to an old metal-rich stellar population.
The main difference  with respect to ASK~0 is the continuum, which steepens
redward of 6000~\AA\  (Fig.~\ref{dust_effect}),  and is a signature of dust 
reddening.
\modified{
Additional independent arguments also corroborate 
that ASK~1 owes much of its red colors to reddening.
ASK~1 galaxies tend to have very elongated morphologies,
a fact difficult to interpret unless they are  edge-on disks
\citep{2011ApJ...735..125S}, which are known
to be significantly dust reddened with respect to
their face-on counterparts
\citep[e.g.,][]{1994AJ....107.2036G,2010MNRAS.404..792M}.
} 

{$\bullet$ ASK~2} is very similar to ASK~0, so does our assignation -- emission
consistent with AGN or LINER-like excitation plus absorption corresponding
to old metal-poor stars. The difference is in the continuum, which is somewhat
redder in ASK~0, and also in the emission line [SIII]$\lambda$9069, which shows up in ASK~2 but
not in ASK~0. 

\begin{figure}
\includegraphics[width=.5\textwidth]{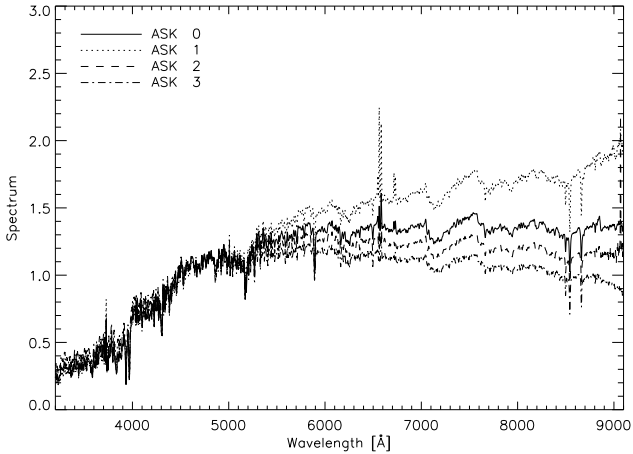}
\caption{
Full spectral range for ASK~0,1, 2, and  3 as indicated.
The absorption features are similar but the 
continua become redder as the ASK number increases.
This change is due to the aging of the stellar populations,
except for ASK~1, which reflects enhanced extinction.
}
\label{early_types}
\end{figure}

{$\bullet$ ASK~3} also shows an absorption line spectrum with weak emission.
Emissions and absorptions are similar to those of ASK~0, therefore,
the associated stellar population is  old. As it happens with ASK~0,
[OIII]$\lambda$5007 and H$\beta$ are too weak to decipher whether
the emission is Seyfert or LINER-like. 
The difference with ASK~0, 1 and 2 is the continuum, which is bluest in ASK~3
(see Fig.~\ref{early_types}).  Except for ASK~1, such variation reflects 
differences in the stellar populations, ASK~3 being the youngest. 
ASK~0, 2, and 3 were used to select a clean  sample of red ellipticals 
by \citet{2012A&A...540A.136A}.

\begin{figure}
\includegraphics[width=.5\textwidth]{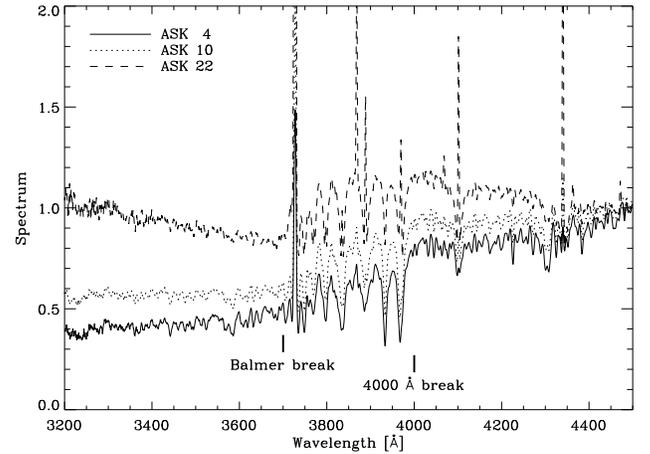}
\caption{
Gradual variation of the breaks at 3650\,\AA\ and 
4000\,\AA\ that show the evolution from 
ASK~4  to ASK~22.
ASK~4 has the oldest stellar population in the series. 
The 4000\,\AA\ break of ASK~4 is more 
intense than the Balmer break. Both are similar in ASK~10, and
the 4000\,\AA\ break has disappeared in the case of ASK~22.
}
\label{breaks}
\end{figure}

{$\bullet$ ASK~4} spectrum has absorption lines and significant
emission lines. The continuum is fairly red, similar to that of ASK~0 in 
Fig.~\ref{dust_effect}. ${\rm H}\alpha \simeq 2\times {\rm [NII]}\lambda6853$
and H$\beta \gtrsim$ [OIII]$\lambda$5007,  therefore, according to
the decision tree (Fig.~\ref{decision}), it should be a LINER-like galaxy. 
However,  it is in the region of the BPT diagram where AGN activity and 
star-formation are difficult to disentangle \citep[see Fig.~13 in][]{2010ApJ...714..487S}.
The absorption line spectrum shows both the Balmer break 
and the 4000\,\AA\ break, which correspond to a mixture of old and young stellar
populations. The region around the break is shown in Fig.~\ref{breaks}.

{$\bullet$ ASK~5} has absorption and emission lines.
The continuum is significantly bluer than that for ASK~4.
 ${\rm H}\alpha \simeq 2\times {\rm [NII]}\lambda6853$
and H$\beta \gtrsim$ [OIII]$\lambda$5007,  therefore, according to
the decision tree (Fig.~\ref{decision}), it should be a LINER-like galaxy (with
the caveat issued for ASK~4 still applying).
The absorption line spectrum shows both the Balmer break at 3650\,\AA\
and the 4000\,\AA\ break, which correspond to a mixture of old and young stellar
populations (like ASK~4 in Fig.~\ref{breaks}).

\begin{figure}
\includegraphics[width=.5\textwidth]{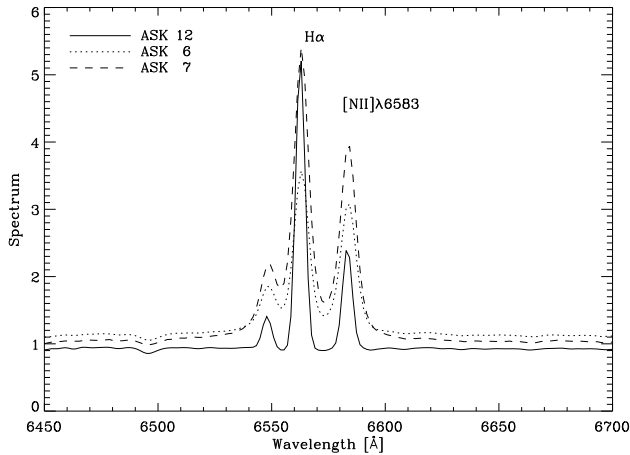}
\caption{
Line width differences between Seyfert galaxies with broad lines 
(ASK~6 and ASK~7), and starburst galaxies with narrow lines (ASK~12).
Even if broad, the lines of ASK~6 and ASK~7 are not broad
enough to be regarded as Seyfert~1.
}
\label{broad_lines}
\end{figure}

{$\bullet$ ASK~6} has intense emission lines on top of an
absorption spectrum. Emission lines are broad (Fig.~\ref{broad_lines}),
but not broad enough  to be a Seyfert~1 galaxy (larger than 2000\,km\,s$^{-1}$).
It appears in the Seyfert region of the BPT diagram, therefore it is a Seyfert 2. 
[FeVII]$\lambda$6087 and [NeV]$\lambda$3426 show up in emission 
confirming the AGN nature of the emission. 
It shows the Balmer break at 3650\,\AA\
and the 4000\,\AA\ break, which correspond to a mixture of old and young stellar
populations. The breaks are extremely similar to that of ASK~4 in Fig.~\ref{breaks}.

{$\bullet$ ASK~7} is similar to ASK~6. The emission lines are broad
 (Fig.~\ref{broad_lines}), and it is also classified as Seyfert 2 according to the 
decision tree. 
[FeVII]$\lambda$6087 is detected, confirming the AGN nature
of the emission (Fig.~\ref{agn_excitation}). 
The absorption spectrum has a clear 4000\,\AA\ break,
and the Balmer break is present but less pronounced than in the case
of  ASK~6
because the Balmer continuum rises blueward of the Balmer break.
The absorption spectrum is also produced by a mixture of old and young stars,
but probably younger than for ASK~6.

{$\bullet$ ASK~8} is similar to ASK~6 and 7, but the emission lines
are even broader. It is also a Seyfert 2.
[FeVII]$\lambda$6087 is detected,
confirming the AGN nature
of the emission (Fig.~\ref{agn_excitation}). 
Following the trend from 
ASK~6 to ASK~7, the absorption spectrum shows the two breaks (Balmer
and 4000\,\AA ), but the Balmer continuum (blueward 
of 3650\,\AA )
is more intense. 
The spectrum is also produced by a mixture of old and young stars,
but probably younger than for ASK~7.

{$\bullet$ ASK~9} has absorption and emission lines.
The lines are narrow, and 
${\rm H}\alpha \simeq 2\times {\rm [NII]}\lambda6853$
with  H$\beta \gtrsim$ [OIII]$\lambda$5007. 
According with the decision tree it has LINER-like emission,
 although it is close to being classified as a metal-rich starburst.
The absorption spectrum presents well defined Balmer and 
4000\,\AA\ breaks, therefore, it is produced by  a mixture of old and young 
stellar populations. The region of the breaks is similar to that of ASK~5.

{$\bullet$ ASK~10} has absorption and emission lines. The continuum
is similar to that of ASK~9 except that it becomes redder beyond 7000\,\AA .
${\rm H}\alpha \simeq 2.5\times {\rm [NII]}\lambda6853$
with  H$\beta \sim 2\times {\rm [OIII]}\lambda$5007. 
According with the decision tree it corresponds to a metal-rich starburst,
 although it is close to the divide with LINER-like emission.
The absorption spectrum is almost identical to the spectrum of ASK~9
in the region of the  Balmer and 4000\,\AA\ breaks, and 
it is produced by  a mixture of old and young  stellar populations.

{$\bullet$ ASK~11} has both absorption and emission, but the 
emission lines are very intense. The red continuum is redder than in 
ASK~9 and 10, but the emission lines of ASK~11 are stronger.
${\rm H}\alpha \simeq 3\times {\rm [NII]}\lambda6853$
with  H$\beta \sim 2\times$ [OIII]$\lambda$5007. 
It corresponds to a metal-rich starburst, although it is close to 
the border to present LINER-like emission.
The absorption spectrum is almost identical to the spectrum of ASK~9
in the region of the  Balmer and 4000\,\AA\ breaks, except 
that the contribution of the Balmer lines is more important.
It is produced by  a mixture of old and young  stellar populations,
but the young population is more important than in the case of ASK~9 and
10.

{$\bullet$ ASK~12} spectrum has both absorption and emission lines.
The continuum is bluer than that of ASK~10 and 11, but the emission lines
are weaker. ${\rm H}\alpha \simeq 2.5\times {\rm [NII]}\lambda6853$
with  H$\beta \sim 1.5\times {\rm [OIII]}\lambda$5007.
It represents a typical metal-rich starburst -- it is right on the 
head of the {\em seagull} of the local BPT diagram corresponding 
to prototypical starbursts.
The absorption line spectrum has  the Balmer and 4000\,\AA\ breaks,
but the  4000\,\AA\ break is less pronounced than that in 
ASK~10 and 11, and the Balmer series more intense. The spectrum 
corresponds to mixed old and young  stellar populations,
but the young population is more important than in the case of ASK~9, 
10, and 11.

{$\bullet$ ASK~13} spectrum has both absorption and emission lines.
The continuum is bluer than that of ASK~11 and 12, but the emission lines
are weaker. ${\rm H}\alpha \simeq 3\times {\rm [NII]}\lambda6853$
with  H$\beta \sim 2\times$ [OIII]$\lambda$5007. It is a starburst.
The absorption line spectrum has  the Balmer and 4000\,\AA\ breaks,
and they are almost identical to those for ASK~12. The spectrum 
corresponds to mixed old and young  stellar populations similar
to ASK~12.

{$\bullet$ ASK~14} spectrum has both absorption and emission lines.
The continuum is bluer than that of ASK~12 and 13, and the  emission lines
more pronounced.
${\rm H}\alpha \simeq 3\times {\rm [NII]}\lambda6853$
with  H$\beta \sim 1.5\times {\rm [OIII]}\lambda$5007.
It corresponds to a typical metal-rich starburst.
The absorption line spectrum has  the Balmer and 4000\,\AA\ breaks,
but the  4000\,\AA\ break is barely noticeable. The spectrum 
corresponds to mixed old and young  stellar populations,
but the young population is more important than in the case of ASK~12, 
and 13.

\begin{figure}
\includegraphics[width=.5\textwidth]{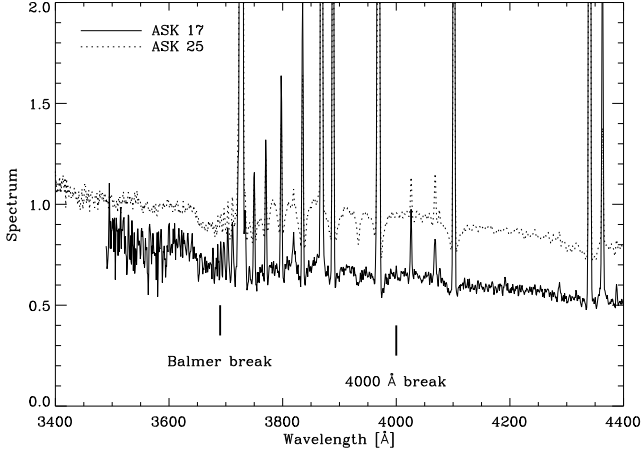}
\caption{
Spectra of HII galaxies. ASK~25 presents no 4000\,\AA\ break, but the 
Balmer break is clear, and the Balmer continuum appears in emission.
The Balmer series shows both emission and absorption.
ASK~17 represents a more extreme case where absorptions are
almost absent, even those of the Balmer series.
Note that the emission lines are well out-of-scale,
with peaks of the order of 15 times the continuum. 
}
\label{hiigalax}
\end{figure}

{$\bullet$ ASK~15} is a pure emission line spectrum. The EW of H$\beta$
is of the order 200\,\AA\ therefore, according the decision tree, it is 
an HII galaxy.
${\rm H}\alpha \gg  {\rm [NII]}\lambda6853$
with  H$\beta \ll  {\rm [OIII]}\lambda$5007, which corresponds to
a low-metallicty starburst.
The spectrum shows 
neither the Balmer break nor the 4000\,\AA\ break
(even more extreme than ASK~17 in Fig.~\ref{hiigalax}). 
There are no metallic lines, and 
even the Balmer series shows no trace of  absorption. This 
spectrum corresponds to the youngest stellar populations 
of the ASK series.  ASK~15 has only 68 members \citep{2010ApJ...714..487S},
most of which look compact galaxies, like those described
by \citet{2009MNRAS.399.1191C} and \citet{2010ApJ...715L.128A}, but a 
few of them are HII regions in resolved galaxies.

{$\bullet$ ASK~16} spectrum has both absorption and emission lines.
The continuum is bluer than that of ASK~13 and 14 
including the upturn at the UV.
${\rm H}\alpha \gg {\rm [NII]}\lambda6853$
with  H$\beta \sim$ [OIII]$\lambda$5007, which
corresponds to a metal-poor starburst.
The absorption line spectrum does show the Balmer break,
but the 4000\,\AA\ break is barely noticeable. 
Consequently,
the absorption line spectrum corresponds to a young  stellar population,
with hints of an old component.
\modified{
The TiO bands are hardly noticeable.
}

{$\bullet$ ASK~17} spectrum has only emission lines.
(One can barely notice the absorption of some of the 
Balmer lines; see Fig.~\ref{hiigalax}). Since the EW of H$\beta\simeq$ 150\,\AA ,
according to the decision tree  ASK~17 is an HII galaxy. 
The continuum is as blue as that of  ASK~15 and includes 
the UV upturn (Fig.~\ref{hiigalax}).
${\rm H}\alpha \gg {\rm [NII]}\lambda6853$
with  H$\beta \ll $ [OIII]$\lambda$5007, which
corresponds to a metal-poor starburst.
Even though absorption lines are not obvious, the spectrum shows the Balmer
break (see Fig.~\ref{hiigalax}). This two features correspond to 
extremely young stellar populations (although not as young as those 
involved in ASK~15).

{$\bullet$ ASK~18} spectrum has absorption and strong
emission lines. The continuum is similar to that of ASK~16 but the 
upturn of the UV continuum is more pronounced. The emission lines
are also stronger than those of ASK~16.
${\rm H}\alpha \gg {\rm [NII]}\lambda6853$
with  H$\beta \simeq $ [OIII]$\lambda$5007, which
corresponds to a metal-poor starburst.
The absorption line spectrum does show the Balmer break,
but it does not have a 4000\,\AA\ break (similar to ASK~22 in Fig.~\ref{breaks}). 
The absorption line spectrum corresponds to a young  stellar population.

{$\bullet$ ASK~19} spectrum presents absorption and strong
emission lines. The continuum is bluer than that of ASK~18 but 
the emission lines are weaker. 
${\rm H}\alpha \gg {\rm [NII]}\lambda6853$
with  H$\beta \ll $ [OIII]$\lambda$5007, which 
corresponds to a metal-poor starburst.
The absorption line spectrum shows the Balmer break,
but it does not have a 4000\,\AA\ break, and it is similar to ASK~18. 
It corresponds to a young  stellar population.

{$\bullet$ ASK~20} spectrum is dominated by strong emission lines
but it also shows weak absorptions in the Balmer lines. 
The weak continuum is as blue as that of  ASK~15, 17 or 19, and includes an 
UV upturn. 
We identify ASK~20 as an  HII galaxy.
${\rm H}\alpha \gg {\rm [NII]}\lambda6853$
with  H$\beta \ll {\rm [OIII]}\lambda$5007, which
corresponds to a metal-poor starburst.
The absorption line spectrum does show the Balmer break,
but it does not have a 4000\,\AA\ break. It also contains metallic lines
(Ca\,II~H\,and\,K).
The absorption line spectrum corresponds to a young  stellar population.
Stars are older than those in ASK~17, but younger than those in ASK~19.
\modified{
The TiO bands are absent, and the IR Ca triplet is almost gone with
a hint of showing up in emission.
}

{$\bullet$ ASK~21} is very similar to ASK~20, except that the 
lines are somewhat weaker.  Probably the gas-phase metallicity is a bit higher in 
ASK~21 as judged from the ratio between  ${\rm H}\alpha$ and ${\rm [NII]}\lambda6853$.
In any case, the starburst is metal-poor.

{$\bullet$ ASK~22} spectrum presents absorption and strong
emission lines. The continuum is similar to ASK~19, but bluer.
The emission lines are somewhat stronger than those in ASK~19. 
${\rm H}\alpha \gg {\rm [NII]}\lambda6853$
with  H$\beta \ll {\rm [OIII]}\lambda$5007, which 
corresponds to a metal-poor starburst. The gas metallicity is a bit higher in 
ASK~19, as judged from the ratio between  
${\rm H}\alpha$ and ${\rm [NII]}\lambda6853$.
The absorption line spectrum shows the Balmer break,
but it does not have a 4000\,\AA\ break, and it is similar to ASK~19. 
The absorption line spectrum corresponds to a young  stellar population,
probably  younger than that in ASK~19.

{$\bullet$ ASK~23} spectrum is very similar to that of ASK~22 and ASK~19,
except for  having
larger emission lines. The continuum is also a bit bluer.
It corresponds to a metal-poor starburst with a young stellar population,
presumedly  younger than that for ASK~19 and 22.

{$\bullet$ ASK~24} has a spectrum similar to ASK~23 (and so to ASK~22 and 19),
with stronger emission lines. The continuum is bluer than in ASK~23. 
As judged from the ratio between ${\rm H}\alpha$ and ${\rm [NII]}\lambda6853$,
the gas metallicity of ASK~23 is higher. 

{$\bullet$ ASK~25} spectrum is similar to ASK~20 and 21, with 
the continuum a bit redder, and the lines weaker.
As judged from the ratio between ${\rm H}\alpha$ and ${\rm [NII]}\lambda6853$,
the gas metallicity of ASK~20 and 21 are smaller. 
The Balmer continuum shows up in emission (Fig.~\ref{hiigalax}).

{$\bullet$ ASK~26} spectrum is similar to ASK~25, with 
the continuum a bit redder, and the lines weaker.
As judged from the ratio between ${\rm H}\alpha$ and ${\rm [NII]}\lambda6853$,
the gas metallicity of ASK~25 is smaller. 

{$\bullet$ ASK~27} spectrum is very similar to that of ASK~25. 
\smallskip

In order to carry out the comparison of this qualitative analysis with the quantitative analysis 
in Sect.~\ref{starlight}, we define a 
{\em stellar age index} that sorts the ASK classes according
to the age of their stellar populations.
The stellar age index (SAI) is defined as follows. Based 
on the absorption lines in the region containing the UV breaks and the 
UV continuum, we order the ASK classes according to their relative stellar 
ages. For instance, having stronger broader Balmer lines implies being older.
Since the ordering is rough, we allow for several classes to share the same age bin. 
Once the order has been set,
we assign a sequential number to this order from the older stellar
population SAI=0 (ASK~0) to the youngest stellar population SAI=13 (ASK~20).
The SAI thus defined qualitatively orders the stellar populations from 
the oldest to the youngest,  although it does not assign specific ages to the 
ASK classes. SAIs  are included in Table~\ref{table_summary}.

Similarly, a gas metallicity index (GMI) is defined
to sort the classes according to their gas-phase 
metallicities. In this case we use the N2 index, 
i.e.,  $\log([{\rm NII}]\lambda6583/{\rm H}\alpha)$,
which is a well known proxy for gas metallicity
(see item~\#~\ref{bptitem} in Sect.~\ref{list_features}).  
Again, the index (i.e., N2 renamed as GMI) is 
listed among the properties of the classes
in Table~\ref{table_summary}.
GMI is used for comparison with the quantitative analysis described
in Sect.~\ref{quantitative_lines}.

\section{Quantitative analysis of the stellar populations using Starlight}\label{starlight}

We use the star formation histories (SFH) derived 
using the code {\sc starlight} \citep{2005MNRAS.358..363C,2007MNRAS.381..263A} 
to cross-check the qualitative analysis carried out 
in the previous sections.
{\sc starlight} decomposes the observed absorption spectrum in terms of a 
sum of single stellar populations (SSP), i.e., coeval starbursts with an
assumed initial mass function, a common metallicity ({\em the metallicity}),  
and observed with a time-lag with respect to the burst ({\em the age}). 
Each SSP produces a known spectrum, and a linear combination of these
spectra is fitted to the observed spectrum, being the amplitudes
applied to each SSP the  free parameters of the fit.
\modified{
Extinction is modeled as a foreground dust screen, with its wavelength 
dependence given by the extinction law of \citet{1994ApJ...429..582C},
and then scaled during the fitting process using a single 
degree of freedom.
(Other extinction laws were tried and yield equivalent results; see
\citeauthor{2007MNRAS.381..263A}
\citeyear{2007MNRAS.381..263A}.)
The amplitudes of the SSPs represent
} 
the measured SFH.
{\sc starlight} uses the Metropolis scheme to carry out the $\chi^2$
minimization  
\citep[for a full description of the code, see][]{2005MNRAS.358..363C}.
The amplitudes thus derived are proportional to the fraction
of the galaxy mass produced by  each individual SSP burst,
and they are the parameters used in our study (after suitable
normalization to 100).  In our particular rendering, 
{\sc starlight}
employs 150 SSPs from \citet{2003MNRAS.344.1000B}, combined
according to the  Padova~1994 evolutionary tracks 
\citep[][and references therein]{1996A&AS..117..113G}.
The SSPs cover a grid of 6 metallicities (from 0.005 to 2.5 times solar)
and 25 ages (from 1 Myr to 18 Gyr). Further details are given
in Sect. 2.1 of \citet{2007MNRAS.381..263A}.
Under these assumptions, we computed the SFH
of each galaxy with a spectrum in SDSS-DR7.

\begin{figure*}
\includegraphics[width=0.9\textwidth]{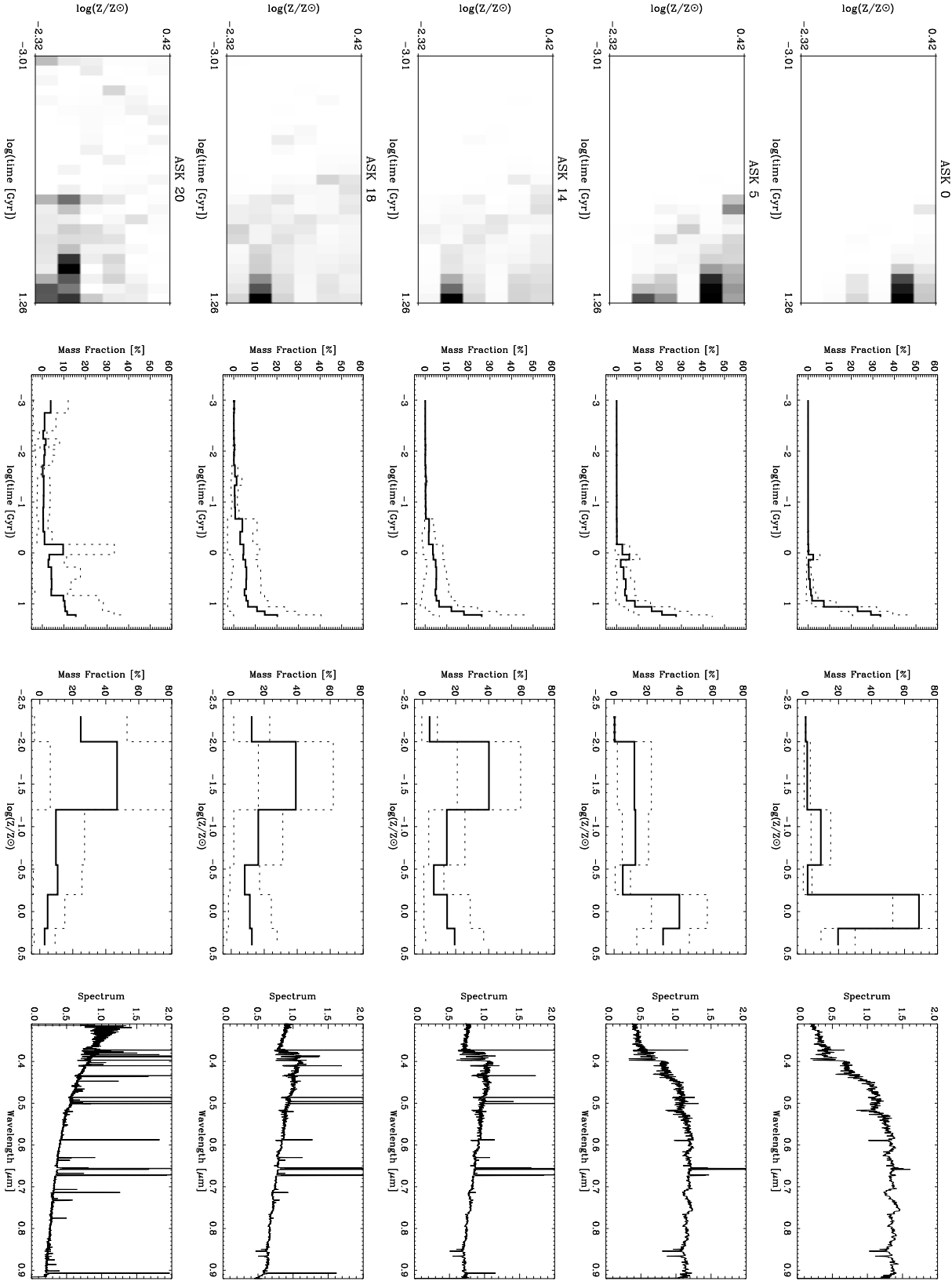}
\caption{
\tiny
(The figure must be rotated by $90^\circ$.)
Each row corresponds to a particular ASK class as indicated.
First column: 
SFHs derived from the application of 
{\sc starlight} to spectra of individual galaxies,
which are then averaged according to their classes.
The classes on display have been
chosen so that they cover the full range of possibilities, from the oldest reddest
stellar populations (ASK~0) to the youngest bluest ones (ASK~20).
ASK~5, 14  and 18 represent intermediate cases.
The SFHs are two dimensional functions with the abscissae representing look-back time,
and the ordinates metallicity. The scale of grays goes from maximum to minimum.  
Second column: the solid lines show the average of the SFHs along the
metallicity axes.
The dotted lines correspond to $\pm$~one standard deviation
\modified{
from the average
}
considering all the 
galaxies included in a given ASK class.
Third column: same as the second column except that 
the average of the SFHs is carried out along the time axes.
The spectrum of each ASK class is also shown for reference in the fourth column.
}
\label{sfh_ask}
\end{figure*}

\begin{figure*}
\includegraphics[width=0.9\textwidth]{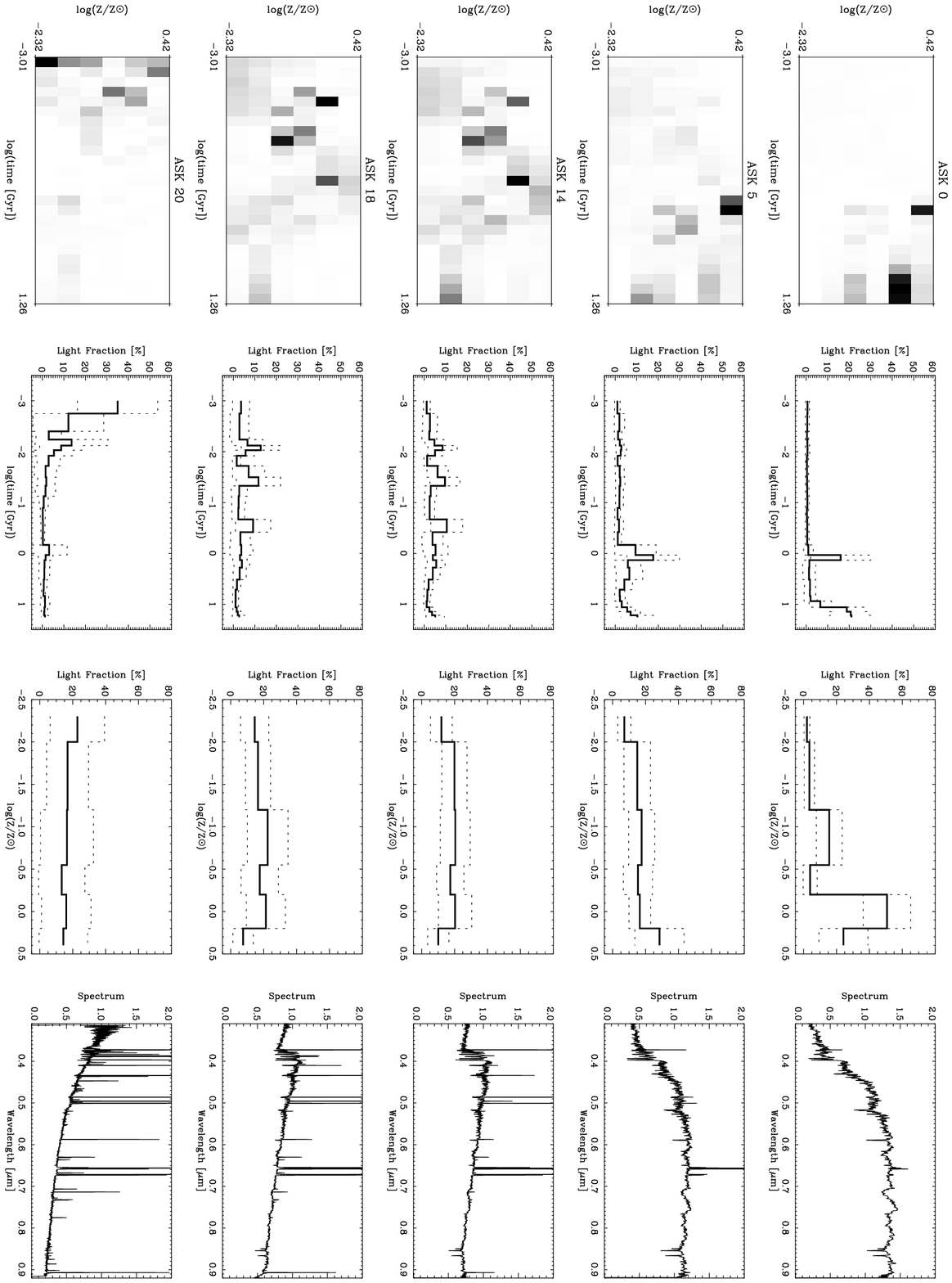}
\caption{
(The figure must be rotated by 90$^\circ$.)
 Percentage of light corresponding to each one
of the SSPs (first column), and its projection in the age axis (second column), 
and in the metallicity axis (third column). Same as 
Fig.~\ref{sfh_ask}, except that the mass of each
component has been weighted by the corresponding light-to-mass ratio.
The ASK classes have been chosen to represent the full range of possibilities,
and they
are the same as those in Fig.~\ref{sfh_ask}.
See the caption of Fig.~\ref{sfh_ask} for further details.
}
\label{sfh_ask_light}
\end{figure*}

Figure~\ref{sfh_ask} shows mean SFHs 
for a number of representative ASK classes. 
The average considers all the SDSS-DR7 galaxies in each ASK class.
The classes have been
chosen so that they cover the full range of possibilities, from the oldest reddest
stellar populations (ASK~0) to the youngest bluest ones (ASK~20).
ASK~5, 14 and 18 illustrate intermediate cases.
Figure~\ref{sfh_ask_light} is equivalent to Fig.~\ref{sfh_ask} except
that,  rather than mass, it shows
the percentage of present light (at 4020\,\AA ) 
produced by each one  of the SSPs. Note how light is strongly biased towards 
young populations, as compared to mass which is 
held by old populations. The dotted lines in the figure represent
$\pm$~one standard deviation considering all the galaxies in SDSS-DR7
corresponding to a given ASK class.
These are the histograms used to compute the luminosity-weighted averages 
and dispersions discussed in the next paragraph.

\begin{figure}
\includegraphics[width=0.5\textwidth]{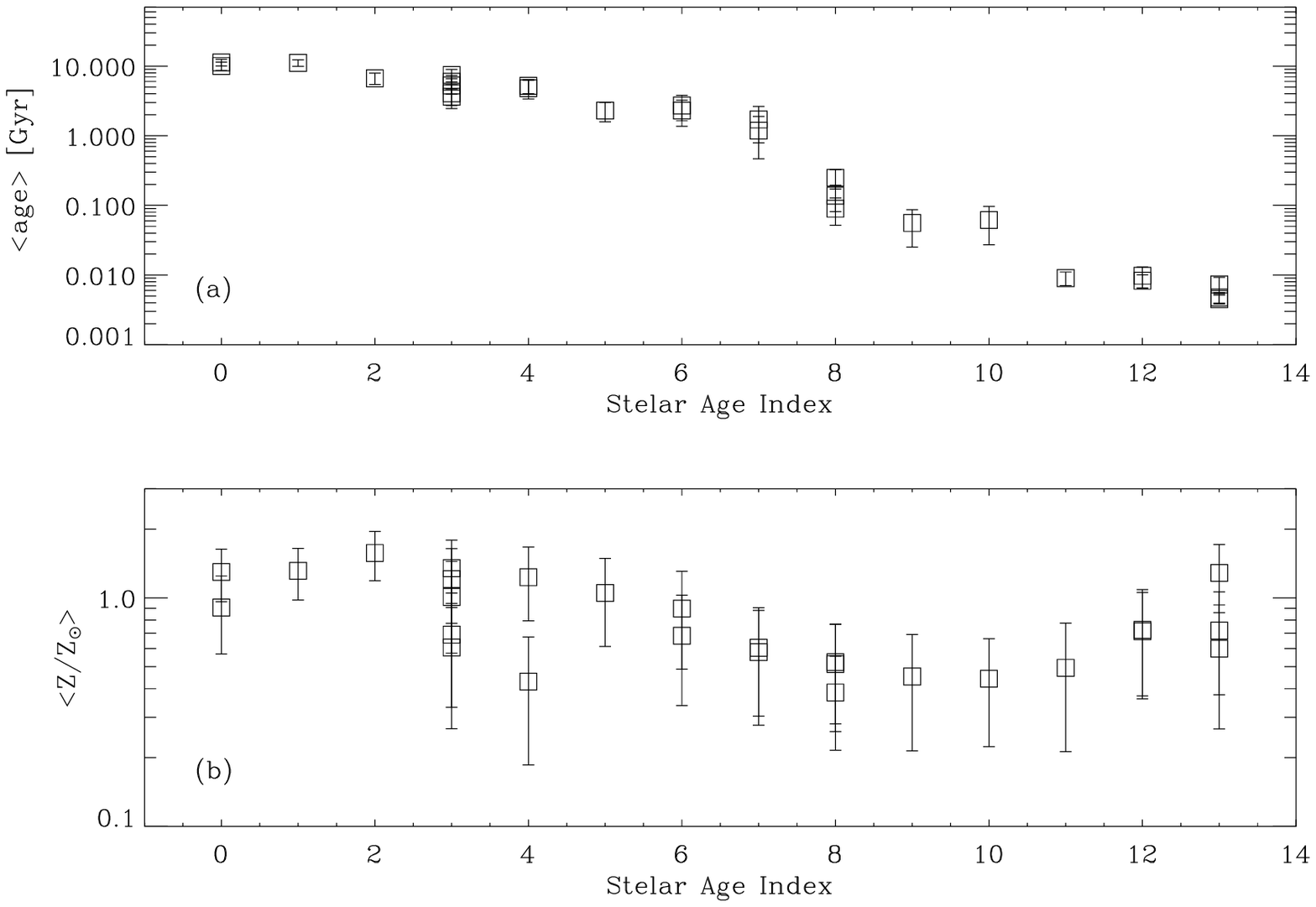}
\caption{
(a) Mean light-weighted age as derived from {\sc starlight} vs SAI (stellar age index). Note the tight
correlation.
(b) Mean metallicity $Z$ derived from {\sc starlight} vs SAI.
The error bars give the rms fluctuations among the ages and 
metallicities of the SSPs included in each class. The scatter of metallicities
is large except for SAI$\leq 2.$ As usual, $Z_\odot$ stands for the
solar metallicity.
}
\label{age_vs_age}
\end{figure}

Figure~\ref{age_vs_age}a shows the relationship between 
the mean luminosity-weighted age as derived from {\sc starlight} 
and the estimate of relative age 
carried out in Sect.~\ref{qualitative_classes} (SAI).
The error bars give the rms fluctuations among the ages of the
SSPs that contribute to each class. 
The correlation age-SAI is extremely good, implying that our quick 
qualitative estimate is consistent with the detailed up-to-date modeling.
Moreover, the existence of an almost one-to-one correlation 
provides specific timescales to our qualitative dating.  
SAI between 0 and 2 correspond to a single old metal rich
population, with ages between 11.2 and 6.7\,Gyr (see the SFH for ASK~0 in 
Fig~\ref{sfh_ask}).
SAI between 3 and 7 has two stellar populations assigned, one old and one young
(Table~\ref{table_summary}). They have mean ages between $7.4\,$Gyr and 1.2\,Gyr.
Finally, from SAI 8  onwards,
we qualitatively find young populations,
and their mean {\sc starlight} ages go from 250\,Myr to 5\,Myr. 

Figure~\ref{age_vs_age}b displays the mean stellar metallicity 
corresponding  to each SAI. The metallicity is high (slightly 
super-solar) when SAI $\leq 2$,
i.e., in the classes our qualitative analysis catalogued as having
old stellar populations. In this case the scatter is fairly 
small (see the error bars in the figure), meaning that all their old
stars are metal-rich. The scatter increases and the mean metallicity 
decreases for younger populations. We interpret this result as
an increase of the number of stellar populations that contribute
to the galaxy spectra, which is corroborated by the SFHs of ASK~5, 14 and
18 in Fig.~\ref{sfh_ask} (with SAI~3, 7 and 8, respectively). 
The stellar metallicity grows slightly for spectra
corresponding to even younger stellar populations, and it becomes slightly 
sub-solar for the youngest ASK classes. The scatter remains
large,  also reflecting the significant
number of stellar components in these galaxies.  

\section{Quantitative analyses of the emission line spectra}
\label{quantitative_lines}

One of the most sophisticated techniques of analysis of ionized nebulae
involves measuring emission-line fluxes of many atomic species 
to derive their relative abundances. 
Adding up all the ionization states of an element provides its abundance.
This  approach is the so-called direct method or 
temperature-based method.
The fluxes depend on atomic parameters
as well as on the physical conditions of the plasma
\citep[e.g.,][]{1981ARA&A..19...77P,1989agna.book.....O,1990ARA&A..28..525S,2004cmpe.conf..115S}.
Once the atomic parameters are known (or assumed),
one can use the observed lines to retrieve, simultaneously, the 
elemental abundances and  the physical conditions
of the nebula. For instance, using collisional excited
lines of the same species having different excitation potentials,
one can determine the electron temperature (e.g., [OIII]$\lambda$4363 and
[OIII]$\lambda$5007). Similarly,  lines of the same species 
with the same excitation potential but different collisional de-excitation
rates, provide diagnostics for the electron density (e.g., [SII]$\lambda$6731 to 
[SII]$\lambda$6717).
We have applied this technique to determine the oxygen abundance 
characteristic of the emission lines of the ASK 
classes that are starbursts.
The actual recipe is described by  
\citet{2003MNRAS.346..105P} and \citet{2008MNRAS.383..209H}, 
and it has been widely used
\citep[e.g.,][]{2010ApJ...715L.128A,2012ApJ...749..185A}.
We refer to the original references for details on the
technique and atomic parameters. 
Whenever possible, the electron temperature
was inferred from [OIII]$\lambda$4363. 
This line weakens with increasing metallicity, therefore,
it cannot be used with the classes of large metallicities
(see item~\#~\ref{line4363} in Sect.~\ref{list_features}).
The problem was bypassed in these cases using 
[SIII]$\lambda$6312 and [SIII]$\lambda$9069 to derive 
the sulphur electron temperature, which was then used
for oxygen after scaling \citep{2006MNRAS.372..293H}.  
ASK classes
17, 20, 21, 22 24, 25, 26, and 27 have [OIII]$\lambda$4363
intense enough  to determine electron temperatures.
The line is not detectable in classes 12, 14, 16, 18, 19 and 23,  
however they show  [SIII]$\lambda$6312,
which we used for deriving electron temperatures.
Finally, classes ASK~10, 11 and 13 do not allow us to measure
either   [OIII]$\lambda$4363 or  [SIII]$\lambda$6312,
and so we could not assign an oxygen abundance 
using the direct method. 
Classes 20, 21, 22, 24, 25, 26, and 27 allow to determine
electron temperatures from both   
 [OIII]$\lambda$4363 and [SIII]$\lambda$6312.
The oxygen abundances obtained using the two ways
of estimating temperature agree within
$\pm$0.02\, dex. All the abundances thus obtained are 
listed in Table~\ref{table_summary}.

\begin{figure}
\includegraphics[width=.5\textwidth]{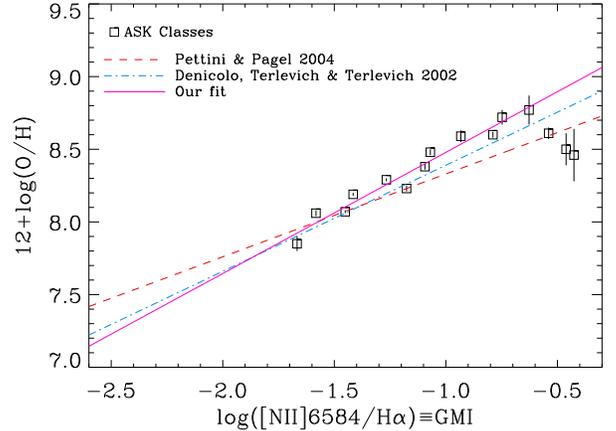}
\caption{
Oxygen abundance of the emission-line component of the templates.
They have been computed as precisely as possible through
estimation of electron temperatures. The oxygen abundance is
represented  versus 
the ratio [NII]$\lambda$6583  to H$\alpha$, which is the 
proxy used to estimate the metallicity
in our qualitative scheme.
Error bars are computed in a Monte-Carlo simulation to be consistent
with the errors assigned to the observed fluxes. 
The straight lines correspond to various estimates of the relationship from
individual galaxies and HII regions by 
\citet[][the dotted-dashed line]{2002MNRAS.330...69D}
and by \citet[][the dashed line]{2004MNRAS.348L..59P},
and from our ASK templates (the solid line).
Our linear fit excludes the three rightmost points, and it
reads,
$12+\log({\rm O/H})=(0.83\pm 0.05)\times\log({\rm [NII]}\lambda6583/{\rm H}\alpha) + (9.31\pm 0.07)$.
}
\label{pplike}
\end{figure}

As we explain in Sect.~\ref{qualitative_classes}, the metallicity of 
the  gaseous component of the template spectra was judged
based on the  ratio  between [NII]$\lambda$6583 and  H$\alpha$. 
This ratio is therefore our qualitative metallicity index
(Table~\ref{table_summary}), which
is compared with the  direct oxygen abundance in 
Fig.~\ref{pplike}. The correlation is 
extremely good, at least from solar metallicity 
(log[O/H]$_\odot = 8.69\pm 0.05$; 
\citeauthor{2009ARA&A..47..481A}~\citeyear{2009ARA&A..47..481A})
to one tenth the solar value. The fluctuations of the
actual data with respect to a linear fit
are just 0.06\,dex, which is 
significantly smaller than the same 
correlation obtained from individual galaxies
-- e.g.,
\citet[][]{2004MNRAS.348L..59P}
claim 0.2\,dex.
From the very good correlation between oxygen abundance and 
[NII]$\lambda$6583/H$\alpha$
we conclude that the qualitative analysis of nebular
metallicities is consistent with the quantitative estimate
using the best techniques available.

\section{Additional results and discussions}\label{additional_results}

\begin{figure}
\includegraphics[width=.5\textwidth]{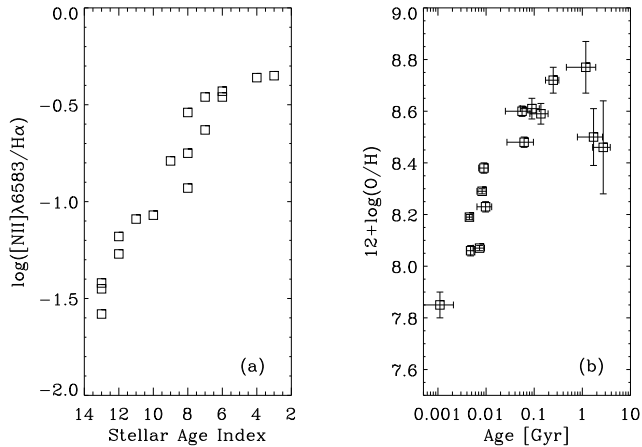}
\caption{
(a) Scatter plot  of the 
index used to determine the gas metallicity, 
$\log{\rm [NII]}\lambda6583/{\rm H}\alpha$,
vs the index used to characterize the age 
of the stellar populations, SAI. They are correlated.
(b) Same representation as (a) but 
using quantitative determinations of gas metallicity
and age. 
The age of the point corresponding to ASK~15 (i.e., the youngest
class with the lowest oxygen content) is just an upper limit, which we 
include to show that the  relationship continues to the youngest
targets.
}
\label{scatterplot}
\end{figure}

\begin{figure}
\includegraphics[width=.5\textwidth]{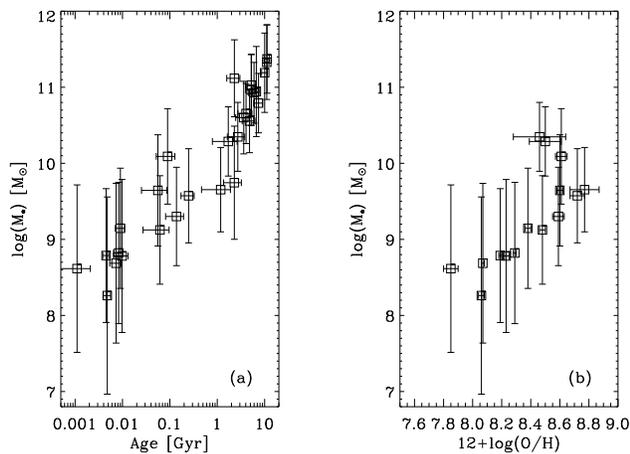}
\caption{
(a) Scatter plot of the mass of the galaxies in each ASK class versus
the mean stellar age. The vertical error bars represent the
standard deviation among all the individual galaxies in each class.
The horizontal error bars are the same as in Fig.~\ref{scatterplot}b.
(b)
Scatter plot  of the mass of the galaxies in a class versus
the gas-phase metallicity. The error bars have the same meaning
as in panel (a).
}
\label{mass_age}
\end{figure}

Figure~\ref{scatterplot}a shows the  index 
used to determine the gas metallicity, 
$\log{\rm [NII]}\lambda6583/{\rm H}\alpha$,
vs the index used to characterize the age 
of the stellar populations, SAI. It is clear that the 
two indices  are correlated, indicating that   the
templates with the lowest oxygen content also
have the youngest stellar populations.
This is explicitly shown in Fig.~\ref{scatterplot}b,
which presents the same kind of relationship but
using quantitative determinations of ages and 
gas-phase metallicities.  
(The two last points deviating from  the linear
relationship will be ignored since the trend they
represent is not present in Fig.~\ref{scatterplot}a, and
they 
have particularly weak [OIII]$\lambda$4363 lines, with the 
uncertainties that this entails
-- see  item \#\,\ref{line4363} in
Sect.~\ref{list_features}.)
\modified{
The correlation is similar to that found by 
\citet{2003MNRAS.340...29C}.
}
The physical origin of the relationship is 
unclear. It may be a side-effect of the galaxy mass 
\citep[a phenomenon often referred to as downsizing; see, e.g.,][]{2006MNRAS.372..933N}.  
First, the mass-metallicity relationship implies
that low-mass galaxies are less metallic \citep[e.g.,][]{1989ApJ...347..875S}.
Second, the mass-age relationship \cite[e.g.,][]{2004Natur.428..625H} implies that 
low-mass galaxies also have younger stellar populations.
Finally,  the bluest ASK classes  
contain more dwarf galaxies \citep{2010ApJ...714..487S},
therefore, they are less metallic and with younger stars,
giving as a side-effect  the observed correlation.
Even though this explanation is feasible, 
the relationship between gas-metallicity and stellar-age 
shown in Figs.~\ref{scatterplot} is so clean that it
looks fundamental rather than derived from  the combined effect of two
other relationships. This conjecture is supported by the scatter plots in
Fig.~\ref{mass_age}, that include the two variables
involved in Fig.~\ref{scatterplot}b plus the galaxy mass.
Assigning masses to the ASK templates is not without ambiguity,
since the spectra of the individual galaxies were normalized
before averaging (Sect.~\ref{ask_class}). However,
we computed the mean and standard deviation among
the masses of all the galaxies in each ASK class\footnote{Stellar 
masses derived from integrated
magnitudes using color-dependent 
mass-to-light ratios from \citet{2001ApJ...550..212B}.}, 
and those are the masses assigned to the classes in 
Figs.~\ref{mass_age}.  One can see that the templates
follow a mass-metallicity relationship (Fig.~\ref{mass_age}b)
and a mass-age relationship (Fig.~\ref{mass_age}a),
but both are less tight than the metallicity-age
relationship in Fig.~\ref{scatterplot}b, which seems to be
the primary relationship.
In short, the properties of the gas and the stars are not 
independent but tightly correlated in real galaxies. Galaxy mass 
does not seem to be the only factor driving such correlation.

Galaxy spectra seem to follow a 1D sequence, with a secondary branch
for AGNs  
\citep{1995AJ....110.1071C,2004AJ....128..585Y,2009AJ....138.1365V,2011MNRAS.415.2417A}.  
In other words, an independent parameter (affine parameter) characterizes 
most properties of galaxy spectra, from the red passive ones to those 
actively forming stars. The actual nature of the affine parameter
is unknown, but the results in this paper suggest it to be 
the mean age of the stellar population.  The ASK templates
can be naturally ordered by mean stellar age  (or by SAI, in our parlance),
and the order thus obtained turns out to be extremely similar to the one
obtained using minimal spanning trees by \citet{2011MNRAS.415.2417A}.
The latter represents a non-trivial exercise to find
the location of the templates in the 1637-dimensional space
where the ASK classification was carried out 
(i.e., a space where each galaxy is a point, and the 1637  coordinates 
represent the flux at particular wavelengths). They are organized in a 
1D sequence with the same order given by the luminosity-weighted mean 
stellar age.
We take the agreement between the two orderings as 
a strong suggestion that stellar age  is the 
affine parameter. Note that the emission line spectrum
is prominent in blue galaxies and so it plays a major
role in shaping the galaxy spectra. That fact that the spectrum
of a galaxy is (mostly) dictated by the age of the stellar population
implies that the emission lines and the absorption
lines are not independent.  This is indeed the conclusion reached in the 
previous paragraph  through a totally different argument.


We argued in Sect.~\ref{ask_class} that the ASK classes are representative 
of all local galaxies since they condense the properties of some one 
million galaxies of the local universe. Even though we endorse 
the statement, it  must be clarified. 
ASK templates are representative of the most common
galaxies, however, some important but uncommon galaxies 
are not included.  In particular, the most massive galaxies that dominate
the centers of galaxy clusters (brightest cluster galaxies and cD
galaxies) are not properly described.  
These massive red galaxies 
have old stellar populations so they are classified as
ASK~0 and ASK~2 \citep{2012A&A...540A.136A}.
However,
they represent a small fraction of all the galaxies in these classes,
so that their contribution to the average (template) spectra of ASK~0 and
2 is negligible. The same may happen with 
other kinds of rare objects like BL Lac \citep[e.g.,][]{1976ARA&A..14..173S},  
objects with extreme star formation rates \citep[e.g.,][]{2009MNRAS.399.1191C}, 
extremely metal poor galaxies \citep[e.g.,][]{2011ApJ...743...77M}, and
others. The fact that some objects may escape the simplified 
schematic in Sect.~\ref{decision_tree} do not invalidate the 
analysis -- it will be useful to indicate that these objects are unusual.

The comparison between Figs.~\ref{sfh_ask} and \ref{sfh_ask_light} evidences 
a fact that is well documented in the literature, but which still 
results somewhat surprising. Most galaxies formed a significant fraction
of their stellar mass long ago when the universe was just a few Gyr old,
even those forming stars today
\citep[e.g.,][]{2004Natur.428..625H,2007MNRAS.381..263A,2011Sci...333..178D}. 
This fact is obviously true for ASK classes
representing passively evolving red galaxies (see
ASK~0 in Fig.~\ref{sfh_ask}), but it also holds true for young ASK classes -- 
see the important contribution of old stellar populations to the ASK~20 SFH in  Fig.~\ref{sfh_ask},
even though its luminosity-weighted mean age is just 4.5~Myr  
(Table~\ref{table_summary}).
When the mass contribution
is transformed to light contribution (Fig.~\ref{sfh_ask_light}), 
it becomes clear how newborn stars outshine the older populations, 
that are heavily underrepresented in the composite galaxy spectrum.
If a galaxy happens to undergo a significant starburst, 
spectrum-wise it looks young.

There is a conspicuous difference between the 
old stellar populations present in passively evolving galaxies 
and in star-forming galaxies. The metallicity of the old stars is
high in passive galaxies and very low in starbursts
(compare the SFHs of ASK~0 and ASK~20 in the first 
column of Fig.~\ref{sfh_ask}).
The dominance of old metal rich stellar populations in red
galaxies is well known  \citep[e.g.,][]{2006ARA&A..44..141R}, 
so does the fact that the old stars in dwarf galaxies of the local 
group have extremely low metallicity 
\citep[e.g.,][]{2001ApJ...562..713M,2010ApJ...722.1864M,2011ApJ...730...14H}.   
 

\section{Conclusions}\label{conclusions}

As argued in the introduction, we have 
sophisticated computer codes for inferring the properties of the stellar 
populations  contributing to the observed galaxy spectra. Similarly, 
tools are available for qualitative diagnostics of the physical
properties of the galaxy gas. They have been developed by specialist
groups, and then kindly offered to a much broader community. 
Technicalities often complicate the interpretation
of the results, therefore, there is a natural 
tendency to apply these sophisticated tools in black-box fashion,
which turns out to be quite unsatisfactory for a physical stand point.
One obtains a  detailed  description of the stars and 
gas producing the observed galaxy spectra, but overlooks the reasons 
why the  computer code  has preferred them rather than other alternatives.  
We provide a simple step-by-step guide to qualitative 
interpretation of galaxy spectra. It is not precise,  and has not
been planed as an alternative to the existing tools. However,
it allows a quick-look that yields the main properties of the spectra
in a intuitive fashion. 
This may be of interest in various applications, e.g., 
to provide physical insight when using sophisticated tools, or 
to interpret noisy spectra.  Moreover, 
the results of the qualitative analysis agree with those inferred using 
up-to-date computer codes. 

The step-by-step guide is described in 
Sect.~\ref{decision_tree}, and it has been
summarized as a simple questionary in Fig.~\ref{decision}. 
Emission and absorption lines are analyzed separately,
which give rise to a classification with one entry for the 
gas and another for the stars.
(In real galaxies, however, the 
properties of gas and stars are 
tightly correlated; see Sect.~\ref{additional_results}.) 
The analysis  has been systematically applied to the set of ASK template 
spectra that resulted from the classification of all 
galaxy spectra in SDSS-DR7 (see Sect.~\ref{ask_class}). 
Their physical properties  are summarized in Table~\ref{table_summary}.
With the caveats pointed
out in Sect.~\ref{additional_results}, the ASK classes
represent a comprehensive set of galaxy spectra, that
go all the way from passively evolving red galaxies (e.g., ASK~0) 
to HII galaxies, dominated by massive newborn stars 
having no absorption lines (e.g., ASK 15).  
Since it works for this set, the analysis should work for most 
galaxies. 

The qualitative analysis is found to be in excellent agreement with
quantitative numerical codes. We show how the index for stellar-age 
(SAI) follows an almost one-to-one correlation with the 
mean stellar age assigned by the code {\sc starlight}  (Fig.~\ref{age_vs_age}).
Similarly, we found how the proxy for gas metallicity
is in good agreement with the (oxygen) metallicity inferred 
by applying the direct method  to the emission
lines of the ASK templates (Fig.~\ref{pplike}).

The ASK templates are freely available (see footnote \#~\ref{my_foot}) 
and, together with their physical properties 
listed in Table~\ref{table_summary}, they can be used as benchmarks  
so that any other galaxy spectrum can be analyzed  by reference to them.

\begin{acknowledgements}

Thanks are due to C. Ramos Almeida and  E. P\'erez-Montero for discussions
and help with references.
This work has been funded by the Spanish MICIN project
{\em Estallidos},
AYA~2010-21887-C04-04.          
ET and RT acknowledge also financial support by the 
Mexican Research Council (CONACYT), 
through grants CB-2005-01-49847, 2007-01-84746 and 2008-103365-F.
%
We are members of the Consolider-Ingenio 2010 Program, grant 
MICINN CSD2006-00070: First Science with GTC.
Funding for the SDSS and SDSS-II has been provided by the Alfred P. Sloan
Foundation, the Participating Institutions, the National Science Foundation, the
U.S. Department of Energy, the National Aeronautics and Space Administration,
the Japanese Monbukagakusho, the Max Planck Society, and the Higher Education
Funding Council for England. The SDSS is managed by the Astrophysical Research 
Consortium for the Participating Institutions (for details,
see the SDSS web site at http://www.sdss.org/).
The {\sc starlight} project is supported by the Brazilian agencies 
CNPq, CAPES and FAPESP and by the France-Brazil CAPES/Cofecub program.

{\it Facilities:} \facility{Sloan (DR7, spectra)
}
\end{acknowledgements}

%
%
\bibliographystyle{aa}

%

\end{document}